\documentclass[iop,twocolappendix]{emulateapj}
\usepackage{natbib}
\usepackage{color}
\usepackage{mathptmx}
\usepackage{amsmath}
\usepackage{url}
\usepackage{multirow}
\usepackage{verbatim}
\usepackage[breaklinks=true]{hyperref} 
\usepackage[figuresright]{rotating}
\usepackage[utf8]{inputenc}
\usepackage{afterpage}  
\usepackage{refcount}
\usepackage{threeparttable}
\usepackage{xcolor}
\usepackage{ulem}
\def\lsim{\lower.5ex\hbox{$\; \buildrel < \over \sim \;$}}
\def\gsim{\lower.5ex\hbox{$\; \buildrel > \over \sim \;$}}

\begin{document}

\shorttitle{Dark Matter Deficient Galaxies From High-velocity Galaxy Collisions}
\shortauthors{Eun-jin Shin, Minyong Jung, Goojin Kwon, Ji-hoon Kim, et al.}

\title{Dark Matter Deficient Galaxies Produced via High-velocity Galaxy Collisions in High-resolution Numerical Simulations}

\author{Eun-jin Shin,$^{1, \dagger}$
Minyong Jung,$^{1, \dagger}$
Goojin Kwon,$^{2, \dagger}$
Ji-hoon Kim,$^{1,3}$\thanks{Co: \href{mailto:me@jihoonkim.org}{me@jihoonkim.org}}
Joohyun Lee,$^{1}$
Yongseok Jo,$^{1}$
Boon Kiat Oh$^{1}$}

\affil{$^{1}$Center for Theoretical Physics, Department of Physics and Astronomy, Seoul National University, Seoul 08826, Korea}
\affil{$^{2}$St John’s College, University of Cambridge, Cambridge, CB2 TP1, United Kingdom}
\affil{$^{3}$\url{me@jihoonkim.org}}
\affil{$^{\dagger}$These authors contributed equally to the article by leading the effort to perform and analyze the simulations discussed herein.}

\begin{abstract}
The recent discovery of diffuse dwarf galaxies that are deficient in dark matter appears to challenge the current paradigm of structure formation in our Universe.  
We describe the numerical experiments to determine if the so-called dark matter deficient galaxies (DMDGs) could be produced when two gas-rich, dwarf-sized galaxies collide with a high relative velocity of $\sim 300\,{\rm km\,s^{-1}}$.  
Using idealized high-resolution simulations with both mesh-based and particle-based gravito-hydrodynamics codes, we find that DMDGs can form as high-velocity galaxy collisions separate dark matter from the warm disk gas which subsequently is compressed by shock and tidal interaction to form stars.   
Then using a large simulated universe {\sc IllustrisTNG}, we discover a number of high-velocity galaxy collision events in which DMDGs are expected to form.  
However, we did not find evidence that these types of collisions actually produced DMDGs in the TNG100-1 run.  
We argue that the resolution of the numerical experiment is critical to realize the ``collision-induced'' DMDG formation scenario.  
Our results demonstrate one of many routes in which galaxies could form with unconventional dark matter fractions.  
\end{abstract}


\section{INTRODUCTION}
\label{sec:1}

Invisible matter named ``dark matter'' constitutes a significant portion of our Universe.  
Dark matter, which interacts only gravitationally, is required on top of baryonic matter to hold galaxy clusters together \citep[e.g.,][]{1933AcHPh...6..110Z}, to explain the motion of satellite galaxies \citep[e.g.,][]{1974ApJ...193L...1O,1974Natur.252..111E}, and to account for the high rotation velocity of stars and gas at large galactic radii \citep[e.g.,][]{1970ApJ...159..379R,1975ApJ...201..327R}.
According to the conventional cold dark matter (CDM) model, structure formation occurs in a ``bottom-up'' manner, where small primordial density fluctuations first grow by gravitational instabilities \citep[e.g.,][]{1982ApJ...263L...1P}. 
Small halos then grow into more massive ones over time via mergers and accretions \citep[e.g.,][]{1984Natur.311..517B}. 
Once the dark matter's gravity creates the potential wells, baryonic matter falls into these wells where star formation and feedback processes ensue \citep[e.g.,][]{1978MNRAS.183..341W}. 
In this paradigm, dark matter is an indispensable driver that facilitates structure formation in the Universe, and should comprise a substantial fraction of mass in any galaxy-sized halos. 

Recently, the discovery of two galaxies that are purportedly deficient in dark matter --- NGC1052-DF2 \citep{2018Natur.555..629V} and NGC1052-DF4 \citep{2019ApJ...874L...5V} --- have intrigued observers and confounded theorists.  
These galaxies are categorized as a class of large yet faint dwarf galaxies known as ultra diffuse galaxies \citep[UDGs; with an effective radius $r_{\rm eff}$ of more than 1.5 kpc and surface brightness $\mu(g)$ of more than 24 ${\rm mag \, arcsec}^{-2}$;][]{2015ApJ...798L..45V}.
For example, based on the velocity measurements of its member globular clusters, the dynamical mass of NGC1052-DF2 is determined to be $M_{\rm dyn} \lesssim 3.4 \times 10^8{\,\rm M_{\odot}}$.  
Meanwhile, its stellar mass is already $M_{\star} \sim 2 \times 10^8\,{\,\rm M_{\odot}}$ with $r_{\rm eff} \sim $ 2.2 kpc, leaving little room for dark matter. 
NGC1052-DF2 has old stellar populations of $\sim$ 9 Gyr \citep{Fensch2019A&A...625A..77F}, and is reported to be metal-poor with [M/H] $\sim$ 1.07.
Despite a few studies suggesting that these galaxies may be closer to us than the previously estimated $\sim 20$ Mpc,\footnote{If they were closer, their stellar mass estimates would decrease, resulting in them being typical galaxies \citep[e.g.,][]{2019MNRAS.486.1192T}.} the latest distance measurement to NGC1052-DF4 using the location of the tip of the red giant branch (TRGB) from the deep HST ACS imaging data indicates that both NGC1052-DF2 and NGC1052-DF4 are indeed members of the NGC1052 group \citep{2019arXiv191007529D}.
If the unusual properties of the two galaxies cannot be explained by measurement errors, they may actually be severely deficient in dark matter. 
Another study also argues that more dwarf galaxies in local universe may have little dark matter \citep{2020NatAs...4..246G}. 
They identified 19 dwarf galaxies that are dark matter deficient within $r_{\rm HI}$, the radius at which HI surface density equals to 1$\,\rm M_{\odot}{\rm pc}^{-2}$. 
Interestingly, 14 out of the 19 galaxies are thought to be isolated, free from any galaxy group, which may imply that those galaxies might have not experienced interactions with other massive neighboring galaxies.

If these observations are confirmed to be accurate, the so-called dark matter deficient galaxies (DMDGs) like NGC1052-DF2 and NGC1052-DF4 appear to challenge aspects of our conventional structure formation paradigm in the Universe. 
In other words, in the contemporary structure formation model, it is unlikely that galaxies that are rich in baryon could have formed in the absence of dark matter. 
Therefore, it is imperative to investigate how these galaxies were born with little dark matter, or have lost their dark matter component along the way. 

Indeed, several formation scenarios have been proposed for DMDGs like NGC1052-DF2 and NGC1052-DF4:
{\it (1)} dwarf galaxies having undergone severe tidal stripping of their dark matter \citep[e.g.,][however, see \citealt{muller2019A&A...624L...6M} which argues that signatures of tidal stripping in NGC1052 are insufficient to come to any definitive conclusion about its role]{2018MNRAS.480L.106O}, 
{\it (2)} old tidal dwarf galaxies (TDGs) formed in a gas cloud that was expelled during tidal interactions from a disk with a high baryonic fraction \citep[e.g.,][]{2012ASSP...28..305D, 2014MNRAS.440.1458D, 2018Natur.555..629V},\footnote{\cite{muller2019A&A...624L...6M} have argued that it is hard to relate the tidal streams near NGC1052-DF2 and DF4 to the old TDG scenarios due to the huge age gap between the streams (age of 1 - 3 Gyr) and NGC1052-DF2 \citep[$\sim$9 Gyr;][]{Fensch2019A&A...625A..77F}.}
{\it (3)} dwarf galaxies formed in a gas cloud that was expelled by outflows from a luminous quasar \citep[e.g.,][however, see \citealt{2018Natur.555..629V} which questions how the large size and low surface brightness of NGC1052-DF2 can be explained by this mechanism]{natarajan1998MNRAS.298..577N}, and
{\it (4)} dwarf galaxies produced during the collision of gas-rich galaxies at a high relative velocity \citep{Silk2019MNRAS.488L..24S} which the present article will focus on.

In the scenario proposed by \cite{Silk2019MNRAS.488L..24S}, during a collision of two gas-rich galaxies at a high relative velocity ($\sim 300\,{\rm km\,s^{-1}}$), non-dissipative dark matter halos pass through each other while dissipative baryons interact to form stars.   
This ``mini-Bullet cluster'' type event \citep[resembling the collision of galaxy clusters that separates dark matter from baryons; e.g.,][]{2006ApJ...648L.109C}  may produce one or more dwarf-sized galaxies with little dark matter --- which we hereby call ``collision-induced'' DMDGs.   
The resulting galaxy's properties could naturally match those of NGC1052-DF2 and NGC1052-DF4 by the following mechanism: 
{\it (1)} The effective radius of a DMDG increases as the removal of dark matter reduces the gravitational potential.
{\it (2)} Supersonic turbulence and shear from the high-velocity collision suppress star formation \citep{anathpindika2018MNRAS.474.1277A, lu2019ApJ...872..171L}, but do not fully quench it \citep{2015ApJ...805..158S}.  
{\it (3)} Low star formation rate, together with the expanded effective radius, yields the low surface brightness.
{\it (4)} The extreme high pressure during the high-velocity galaxy collision may have caused the formation of bright globular clusters \citep{elmegreen1997ApJ...480..235E}.

In this paper, using various gravito-hydrodynamics simulations, we investigate if the observed DMDGs could have been produced when gas-rich galaxies collide with a high relative velocity.     
While several groups have studied if DMDGs are found in large cosmological simulations such as {\sc Illustris} and {\sc Eagle} \citep[e.g.,][see Sections \ref{sec:5} for more discussion]{2018arXiv180905938Y, 2019MNRAS.488.3298J, 2019MNRAS.489.2634H, 2019A&A...626A..47H}, the ``collision-induced'' DMDG formation scenario --- which could potentially better explain the properties of NGC1052-DF2 and NGC1052-DF4 --- has never been numerically tested.  
Here, we first test the viability of the collision-induced DMDG formation mechanism (see Section \ref{sec:3} for its definition and further discussion) by carrying out a suite of idealized galaxy collision simulations with both mesh-based and particle-based codes. 
We then explore the collisional parameter space such as the relative velocity, disk angle, and gas fraction, and determine the subset of the space that permits the DMDG formation. 
We also search for the collision-induced DMDGs in a large simulated universe TNG100-1, and examine what factors are required to form DMDGs via galaxy collisions in a numerical simulation. 

This paper is structured as follows. 
In Section \ref{sec:2} we first describe the two types of simulations used in our study. 
Section \ref{sec:3} presents the DMDG formation in the idealized high-resolution galaxy collision simulations using {\sc Enzo} and {\sc Gadget-2}. 
Section \ref{sec:4} discusses how many high-velocity galaxy collisions have occurred in which a DMDG is expected to form in a large simulated universe TNG100-1. 
In Section \ref{sec:5} we present our effort to search for collision-induced DMDGs in the TNG100-1 universe.  
We summarize our findings and conclude the paper in Section \ref{sec:6}.

\begin{table*}
\centering
\caption{Structural properties of a model galaxy used as a progenitor in our fiducial, idealized galaxy collision simulation}
\begin{tabular}{p{3cm}p{5cm}p{3cm}p{3cm}} 
\hline\hline
&Dark matter halo & Stellar disk & Gas disk \\
\hline
Density profile& \cite{1997ApJ...490..493N}  & Exponential & Exponential \\
\hline
Structural properties
& $M_{200} =5.95\times10^{9}{\,\rm M_{\odot}}$, $v_{200} = 35\, {\rm km\,s^{-1}}$,
$J_{200}=1.17\times10^{12}{\,\rm M_{\odot}} \,{\rm kpc\, km\,s^{-1}}$,         
$c=13$
& $M_{\rm d,\star}= 4.21\times10^{8}{\,\rm M_{\odot}}$,
$r_{\rm d} = 1.3 {\,\rm kpc}$
& $M_{\rm d,gas} = 1.68\times10^{9}{\,\rm M_{\odot}}$, $f_{\,\rm d, gas} =0.8$ \\
\hline
Number of particles&10$^{5}$&10$^{4}$&$1.96\times10^{4}$\\
\hline 
Particle mass & $m_{\rm DM}=3.88\times10^{4}{\,\rm M_{\odot}}$ & $m_{\rm \star, IC}=4.21\times10^{4}{\,\rm M_{\odot}}$ & $m_{\rm gas, IC}=8.59\times10^{4}{\,\rm M_{\odot}}$\\
\hline
\end{tabular}
\vspace{1mm}
\tablecomments{See Section \ref{sec:2.1} for more information.}	
\label{tab:galaxy-parameter}
\vspace{3mm}
\end{table*}

\vspace{3mm}
\section{Simulations}\label{sec:2}

In this section, we describe the two types of simulations we have used to investigate the collision-induced formation scenario of DMDGs:  a suite of idealized galaxy collision simulations at high resolution (80 pc; Section \ref{sec:3}), and a large cosmological simulation with relatively low resolution (Sections \ref{sec:4} and \ref{sec:5}).

\subsection{A Suite of Idealized Galaxy Collision Simulations}\label{sec:2.1}

\subsubsection{Simulation Codes {\sc Enzo} and {\sc Gadget-2}, and Baryonic Physics Adopted}

In Section \ref{sec:3}, we simulate galaxy collisions using both the grid-based code {\sc Enzo} and the particle-based code {\sc Gadget-2}. 
Performing the simulation on two flavors of hydrodynamics solvers is to ensure that any observed features are not artifacts of a particular numerical implementation, but phenomena reproducible by each other.  

{\sc Enzo} is an open-source code that employs 3-dimensional Eulerian structured mesh \citep{2014ApJS..211...19B}.\footnote{The website is http://enzo-project.org/.} 
The equations of gas dynamics are solved using the 3rd-order accurate piecewise parabolic method \citep[PPM;][]{1984JCoPh..54..174C}, with the adaptive mesh refinement (AMR) method that refines the grid to a higher resolution when certain criteria are met. 
In this study, we initialize a $64^3$ root grid in a $(1.311\,{\rm Mpc})^{3}$ box, and allow 8 levels of refinement, achieving a finest spatial resolution of 80 pc. 
At each level, a gas cell is divided into eight child cells when the gas mass in the cell exceeds $8.59\times10^{4}{\,\rm M_{\odot}}$ ($= m_{\rm gas, IC}$, a gas particle mass in the initial condition for {\sc Gadget-2}).
On the other hand, the particle-based code {\sc Gadget-2} is a Tree-Particle-Mesh (TPM) Smooth Particle Hydrodynamic (SPH) code that uses a smoothing kernel to estimate hydrodynamic properties of fluids \citep{2005MNRAS.364.1105S}.\footnote{The website is http://www.h-its.org/tap-software-en/gadget-code/ or http://www.mpa-garching.mpg.de/gadget/.}
The Plummer equivalent gravitational softening length is set to $\epsilon_{\rm grav} = $ 80 pc while the minimum hydrodynamic smoothing length is set to $0.2\, \epsilon_{\rm grav}$. 
Both codes have been widely used in galaxy formation and collision simulations, demonstrating their capabilities to contain and resolve all relevant physics previously discussed in the galaxy evolution process \cite[e.g.,][]{2008MNRAS.384..386C, 2011MNRAS.412.1341D, 2009ApJ...694L.123K, 2019ApJ...887..120K}.  

Among those galaxy-scale physics, for both {\sc Enzo} and {\sc Gadget-2} we include a tabulated radiative cooling in an ionization equilibrium provided by the {\sc Grackle} library \citep{2017MNRAS.466.2217S}, along with the metagalactic UV background radiation at $z=0$ \citep{HaardtMadau12}.
To avoid artificial collapse and fragmentation of the gas, the Jeans pressure is employed and set to 
\begin{equation}
\centering 
    P_{\rm Jeans} = \frac{1}{\gamma\pi}N^{2}_{\rm Jeans}G\rho^{2}_{\rm gas}\Delta x^{2},
\end{equation}
where $\gamma=5/3$ is the adiabatic index, $G$ is the gravitational constant, $\rho_{\rm gas}$ is the gas density, $N_{\rm Jeans}=4$ is the controlling parameter, and $\Delta x=80 \,{\rm pc}$ is the spatial resolution \citep{1997ApJ...489L.179T}. 
Star formation occurs when the gas density exceeds the threshold, $n_{\rm H, \,thres} = 10\, {\rm cm}^{-3}$. 
The gas parcel above the density threshold creates a star particle at a rate that follows the local Kennicutt-Schmidt law \citep{1959ApJ...129..243S,1998ApJ...498..541K},
\begin{equation}
        \frac{d\rho_{\star}}{dt} = \frac{\epsilon_{\star} \rho_{\rm gas}}{t_{\rm ff}},
\end{equation}
where $\rho_{\star}$ is the stellar density, $t_{\rm ff} = (3\pi/(32G\rho_{\rm gas}
))^{1/2}$ is the local free-fall time, and $\epsilon_{\star} = 0.01$ is the star formation efficiency. 
A new star particles inject the thermal energy of $10^{51}$ ergs, mass of $14.8{\,\rm M_{\odot}}$ (of which $2.63{\,\rm M_{\odot}}$ is in metals) per every $91{\,\rm M_{\odot}}$ of stellar mass formed, after a delay time of 5 Myr from its creation.  

It should be noted that the numerical resolution (80 pc) and baryonic physics adopted here match those proposed by the {\it AGORA} High-resolution Galaxy Simulations Comparison Project, needed to reliably describe the galaxy-scale physics in a reproducible manner \citep{2014ApJS..210...14K, 2016ApJ...833..202K}.

\subsubsection{Initial Conditions for Idealized Galaxies and Their Collision}

For our study in Section \ref{sec:3}, we design a collision simulation between two gas-rich dwarf-sized galaxies.  
As a fiducial model, we use a system of two identical galaxies, each of which consists of a dark matter halo, a stellar disk and a gas disk of masses $3.88\times10^{9}{\,\rm M_{\odot}}$, $4.21\times10^{8}{\,\rm M_{\odot}}$, and $1.68\times10^{9}{\,\rm M_{\odot}}$, respectively. 
The total mass of each galaxy is thus $M_{200} =5.95\times10^{9}{\,\rm M_{\odot}}$ with gas fraction $f_{\rm gas} = M_{\rm d, gas}/M_{200}= $ 0.28. 
For both {\sc Enzo} and {\sc Gadget-2}, $10^{5}$ dark matter particles (for a halo following the \cite{1996ApJ...462..563N} profile with concentration $c=13$) and $10^{4}$ star particles (for a disk of an exponential profile) are used in each galaxy.
To initialize the gaseous disk in an exponential profile with a scale radius of $r_{\rm d} = 1.3 \,{\rm kpc}$ and a scale height of $z_{\rm d} = 0.13 \,{\rm kpc}$, {\sc Gadget-2} uses $1.96\times10^{4}$ particles while {\sc Enzo} grids the exponential profile onto an adaptive mesh. 
The initial temperature of the gas disk is set to 10$^{4}$ K and initial metallicity to solar metallicity, $Z_{\odot} = 0.02041$. 
As for the gas in the halo, only {\sc Enzo} sets up a very diffuse gas of uniform density $n_{\rm H}=10^{-6}\,{\rm cm}^{-3}$ and temperature $10^{6}\,{\rm K}$.

The galaxies are initialized and then placed on a collision orbit by the {\sc Dice} code \citep{2016ascl.soft07002P} which  sets up 3-dimensional distributions of particles in various structural components by using the Markov Chain Monte Carlo (MCMC) algorithm, and calculates the dynamical equilibrium of each component by integrating the Jeans equation.\footnote{The website is https://bitbucket.org/vperret/dice/.} 
Table \ref{tab:galaxy-parameter} summarizes the components of our model galaxy used in the fiducial collision simulation (see Section \ref{sec:3.2}).  
Table \ref{tab:gadget-parameter} lists all variations in the collision configuration for our parameter study (see Section \ref{sec:3.3}).

\subsection{A Large Simulated Universe TNG100-1}\label{sec:2.2}

In Sections \ref{sec:4} and \ref{sec:5}, we search for collision-induced DMDGs in a large simulated universe, {\sc IllustrisTNG} \citep{Pillepich2018-ou, 2019ComAC...6....2N}.\footnote{The website is https://www.tng-project.org/.} 
It is the latest iteration of the {\sc Illustris} run, a large cosmological box simulated with the moving-mesh code {\sc Arepo} \citep{Springel2010-jo}. 
The updated physics ingredients adopted in {\sc IllustrisTNG} include magneto-hydrodynamics \citep{pakmor2011magnetohydrodynamics, pakmor2013simulations}, metal advection \citep{naiman2018first}, stellar evolution and galactic winds \citep{Pillepich2018-ou}, and supermassive black hole physics  \citep{weinberger2017blackhole, Weinberger2018-sw}, along with the cosmology consistent with the latest Planck data: $\Omega_{\text{m}}= 0.3089, \Omega_{\Lambda} =0.6911, \Omega_\text{b} = 0.0486, \sigma_8 = 0.8159, n_{\text{s}} = 0.9667$, and $h=0.6774$  \citep{ade2016planck}. 

The {\sc IllustrisTNG} project provides three different simulation volumes ---  TNG50, TNG100, TNG300 (numbers denoting approximate box sizes in Mpc) --- each at three different resolutions --- e.g., TNG100-1 (high), -2 (intermediate), and -3 (low). 
For our analysis in Sections \ref{sec:4} and \ref{sec:5}, we employ TNG100-1, a $(106.5\,{\rm comoving \,\,Mpc})^3$ volume containing $1820^3$ dark matter particles and $1820^3$ gas cells, initially with mass resolution $7.5 \times 10^6 \,\rm M_{\odot}$ and $1.4 \times 10^6 \,\rm M_{\odot}$ for dark matter and gas, respectively.  
The Plummer equivalent gravitational softening length for collisionless particles is $\epsilon_{\rm grav}^{z=0} = 740\, {\rm pc}$, and the minimum adaptive gravitational softening length for gas particles $\epsilon_{\rm gas, min} = 185\,\, {\rm comoving \,\,pc} $.\footnote{In the TNG100-1 run, $\epsilon_{\rm grav}$ is set to 1 comoving $h^{-1} {\rm kpc}$ down to $z = 1$, after which it is fixed to $0.5 \,{\rm (physical)} \,h^{-1} {\rm kpc} = $ 740 pc.  $\epsilon_{\rm gas, min}$ is set to $0.25 \epsilon_{\rm grav} =  0.25 \, {\rm comoving} \, h^{-1} {\rm kpc} $ down to $z = 1$, after which it is fixed to $0.25 \times 0.5 \,{\rm (physical)} \,h^{-1} {\rm kpc} = $ 185 pc. Readers should note the difference in resolution between TNG100-1 and the idealized simulations detailed in Section \ref{sec:2.2} (for discussion on how it affects the formation of DMDGs in galaxy collisions, see Section \ref{sec:5.3}).   \label{footnote:tng-res}} 
The project also offers ``subboxes'' that are spatial cutouts at $<$ 80 times finer output intervals (i.e., a few Myr), for studies requiring better temporal resolution. 
We utilize two subboxes in TNG100-1, each in a $(11.1\, {\rm comoving\,\, Mpc})^3$ volume (see Section \ref{sec:5.2}).

Of particular interest to the readers of this article is the star formation physics in {\sc IllustrisTNG}.  
For TNG100-1, the star formation is based on the Kennicutt-Schmidt relation \citep{1998ApJ...498..541K}, and  occurs stochastically assuming the \cite{Chabrier2003PASP..115..763C}  initial mass function with the density threshold $n_{\rm H,\, thres} = 0.13\,{\rm cm}^{-3}$ and the star formation timescale $t_{\rm SF}=2.2\,{\rm Gyr}$ (see \citealt{Springel2003-lt} and \citealt{Pillepich2018-ou} for details).  
The model includes an effective equation of state (EoS) to account for the impact of unresolved physics on a small scale such as supersonic turbulence and thermal conduction \citep{2013MNRAS.436.3031V}.

Lastly, the  TNG100 snapshots are accompanied by the halo catalogues identified with the friends-of-friends
halo finder \citep[{\sc FOF};][]{Davis1985} and the {\sc SubFind} subhalo finder \citep{Springel2001, dolag2009MNRAS.399..497D}. 
They also come with the merger trees built with the {\sc SubLink} algorithm \citep{Rodriguez-Gomez2015}. 

\vspace{3mm}

\section{Dark Matter Deficient Galaxies (DMDGs) Produced In High-velocity Galaxy Collision Simulations} \label{sec:3}
Using the numerical tools introduced in Section \ref{sec:2.1}, we now study how a DMDG could have formed in a high-velocity galaxy collision.  

\subsection{Definition of A Dark Matter Deficient Galaxy (DMDG)}\label{sec:3.1}

We begin this section with a definition of a DMDG to be used in the subsequent analyses, based on the observed characteristics of NGC1052-DF2 and NGC1052-DF4 --- UDGs of dynamical masses $\sim10^{8}{\,\rm M_{\odot}}$ with bright globular clusters, but with little evidence of dark matter \citep{2018Natur.555..629V, 2019ApJ...874L...5V}. 
In the present paper, we define a DMDG as a stellar mass-dominated galaxy with {\it (1)} its dark matter mass being less than 50\% of the total mass, and {\it (2)} the stellar mass of more than $10^7\,\rm M_{\odot}$.  
The galaxy should also {\it (3)} be self-gravitating, and {\it (4)} last for sufficiently long time so that it is observable before e.g., it is tidally disrupted by a neighboring massive galaxy.   

The last condition is particularly important as we consider the survival of the remnants from the galaxy collision.   
We find in the analysis of the TNG100-1 data (Sections \ref{sec:4} and  \ref{sec:5}) that a significant fraction of the high-velocity collision events occur between two satellite galaxies near a massive host galaxy. 
The host galaxy may then tidally strip any galaxies apart in its proximity, or pull them inwards to disintegrate them, making it challenging for any remnant --- a DMDG or otherwise --- to survive.  
This means that a potential DMDG needs to have enough orbital energy to survive in the host galaxy's strong gravitational influence.  

\begin{figure*}
    \centering
    \vspace{3mm}
    \includegraphics[width=0.95\textwidth]{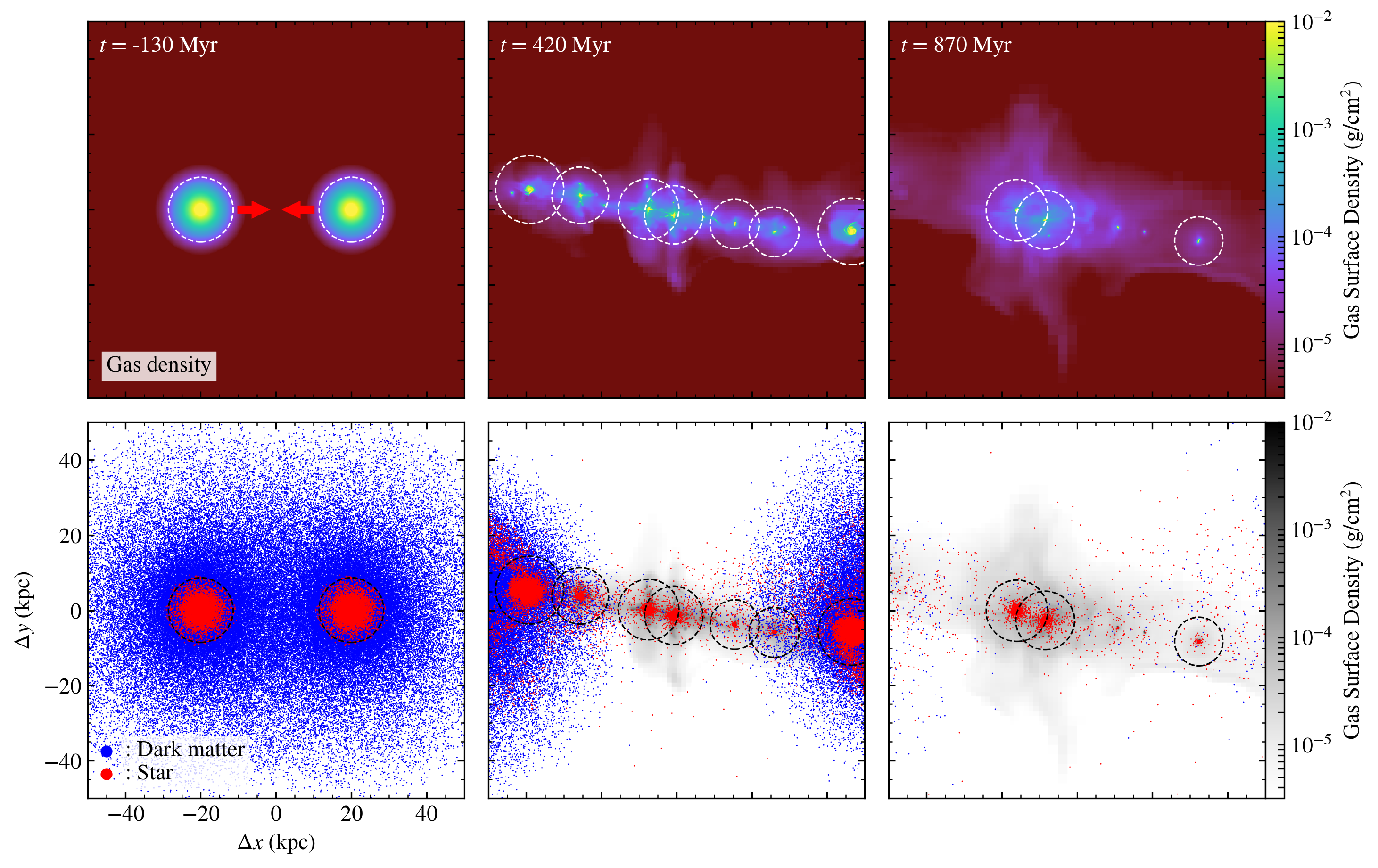}
    \caption{$t = -130, 420, 870$ Myr snapshots of the fiducial, idealized galaxy collision simulation on the grid-based gravito-hydrodynamics code {\sc Enzo} between two gas-rich galaxies of $M_{200} = 5.95\times10^{9}{\,\rm M_{\odot}}$ each.  
    $t$ is the time since the two galaxies' pericentric approach.
    {\it Top}: face-on gas surface density in a 100 kpc box centered on the two galaxies' center of mass which later becomes the point of collision. 
    {\it Bottom}: dark matter particles are colored {\it blue} while newly-created star particles are colored {\it red}.
    The disks of both galaxies lie in the orbital plane ($x$-$y$), approaching each other at $300\,{\rm km\,s^{-1}}$ at $t = -130$ Myr. 
    Galaxies identified by the {\sc Hop} algorithm are marked with {\it dashed circles} of log$_{10}(M_{\star}/{\,\rm M_{\odot}})$ kpc radii.      
    Several dark matter deficient galaxies (DMDGs; almost devoid of {\it blue} particles) have formed as a result of the collision, the most massive one being with $M_{\star,\,<{\rm 5kpc}} = 1.60\times 10^{8}{\,\rm M_{\odot}}$ (see Figure \ref{fig:2}).
    See Section \ref{sec:3.2} for more information.}
\label{fig:1}
\vspace{3mm}
\end{figure*}

\begin{figure*}
    \centering
    \includegraphics[width=1.01\textwidth]{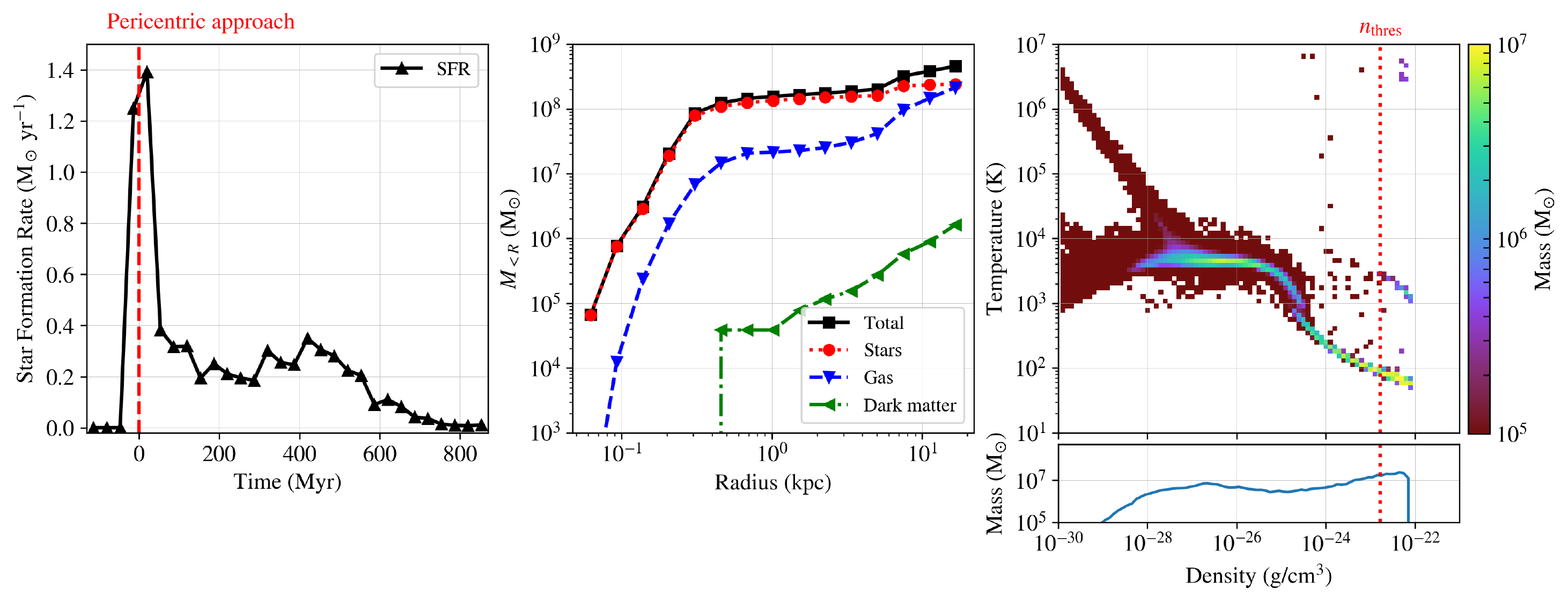}
    \caption{
    {\it Left}: the star formation rate (SFR) estimated for all star particles within 20 kpc from the most massive DMDG born in the {\sc Enzo} simulation shown in Figure \ref{fig:1}.  
    The {\it red vertical line} marks the moment when two galaxies are in a pericentric approach, $t=0$ Myr.
    {\it Middle}:
    the radial profile of the enclosed mass in the most massive DMDG at $t = 870$ Myr, showing that little dark matter exists in the DMDG.
    {\it Right}:
    2-dimensional probability distribution function (PDF) of density and temperature for the gas within 20 kpc from the center of the most massive DMDG at $t =420$ Myr (when the two progenitor galaxies are $\sim$ 100 kpc apart after their first encounter). 
    The {\it red dotted line} denotes the star formation threshold density, $n_{\rm H, \, thres} = 10\, {\rm cm}^{-3}$. 
    Shown in the {\it bottom right} panel is the gas density distribution function.
    The $y$-axis ranges of these panels are kept identical for Figures \ref{fig:2},  \ref{fig:4}, \ref{fig:10} and \ref{fig:12} for easier comparison.
   }
\label{fig:2}
\end{figure*}

\begin{figure*}[h!]
    \centering
   \includegraphics[width=0.95\textwidth]{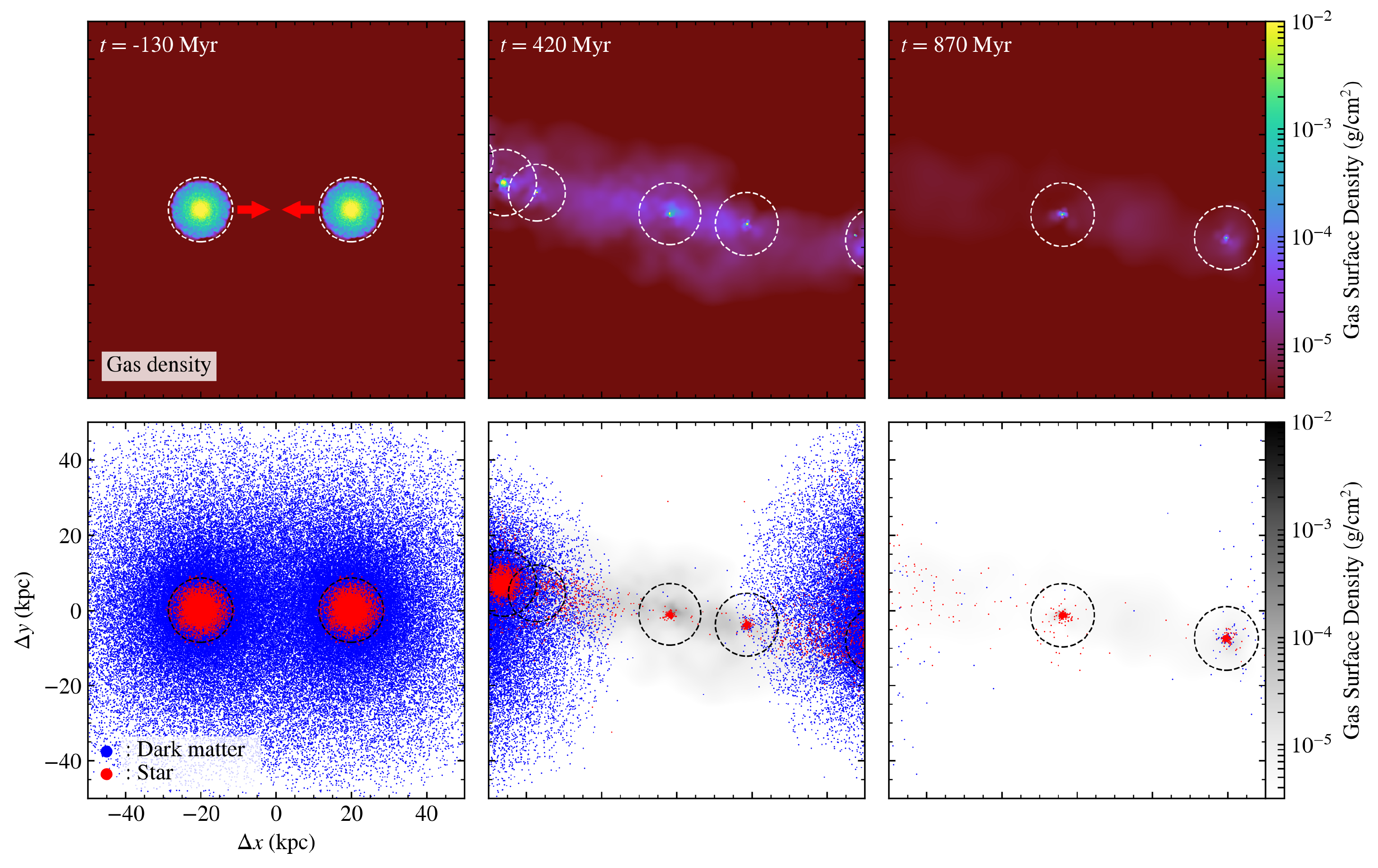}
    \caption{Same as Figure \ref{fig:1}, but depicting the results from a simulation performed on the particle-based gravito-hydrodynamics code {\sc Gadget-2}.
     As in Figure \ref{fig:1}, multiple DMDGs have formed during and after the collision, among which the most massive one is with stellar mass $M_{\star,\,<{\rm 5kpc}} = 2.16\times 10^{8}{\,\rm M_{\odot}}$ (see Figure \ref{fig:4}).
     See Section \ref{sec:3.2} for more information.}
\label{fig:3}
\end{figure*}

\begin{figure*}[h!]
    \centering
    \vspace{-35mm}    
    \includegraphics[width=1.01\textwidth]{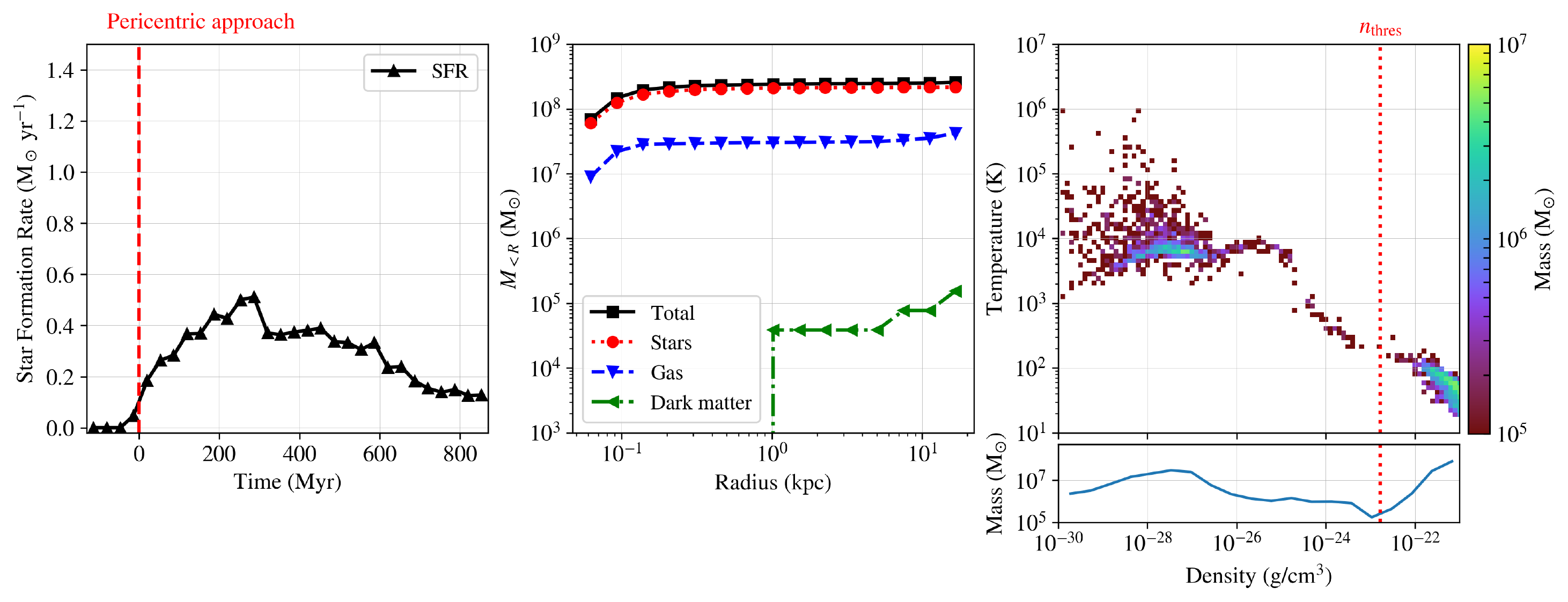}
    \caption{
    Same as Figure \ref{fig:2}, but for the {\sc Gadget-2} run shown in Figure \ref{fig:3}.  
    Comparing with Figure \ref{fig:2}, both simulations indicate that almost all the stars in the resulting DMDG is a direct result of collision-induced star formation ({\it left}), and the DMDG's composition is similarly deficient in dark matter ({\it middle}).  Both runs exhibit cold dense gas above the star formation threshold density $n_{\rm H, \, thres}$, implying continuous collision-induced star formation at $t=420$ Myr ({\it right}).}
    \label{fig:4}
\end{figure*}

\begin{table*}
\centering
\begin{threeparttable}
\caption{A suite of idealized galaxy collision simulations on {\sc Gadget-2} listed with their collision configurations and the properties of the resulting DMDGs}
\vspace{3mm}
\begin{tabular}{@{}ccccccccccc@{}}

\toprule
\multicolumn{5}{c}{Collision parameters}&& Resulting DMDGs& &\multicolumn{3}{c}{Most massive DMDG}\\
\cline{1-5}\cline{7-7}\cline{9-11}
Collision velocity &  Pericentric distance & Disk angle & Mass ratio & Gas fraction  && \# identified &&   $M_{\star}$  & $M_{\rm gas}$ & $M_{\rm DM}$\\
 (${\rm km\,s^{-1}}$)&(kpc)&($^{\circ}$)& &  &  &&&  ($10^8\, \rm M_{\odot}$)  & ($10^8\,\rm M_{\odot}$) & ($10^8\,\rm M_{\odot}$)  \\
 (1)&(2)&(3)&(4)&(5)&&(6)&&(7)&(8)&(9)\\
 \hline
        100  & 1 & 45 & 1:1 &0.28 &&- && -&-&-\\
        100  & 1 & 45 & 1:3 &0.28 &&- && -&-&-\\
        100  & 1 & 45 & 1:10 &0.28 &&- && -&-&-\\
        300  & 1 & 0 & 1:1 &0.28 &&2\tnote{\textcolor{red}{\textdagger}} &&2.16&0.424&0.00155\\
        300  & 1 & 0 & 1:3 &0.28 &&3&&1.28& 0.602& 0.129\\
        300  & 1 & 45 & 1:1 &0.28 &&2\tnote{\textcolor{red}{\textdaggerdbl}} &&0.103&0.699 & 0.00116\\
        300  & 1 & 45 & 1:3 &0.28 &&3&&0.825&0.538& 0\\
        300  & 1 & 45 & 1:10 &0.28 &&- &&-&-&-\\
        300 & 1 & 0 & 1:1 &0.13 &&3&& 0.993&0.586&0\\
        300 & 1 & 0 & 1:1 &0.07 &&2&& 0.303& 0.379& 0\\
         300 & 1 & 0 & 1:1 &0.02 &&- && -&-&-\\
        300 & 3 & 0 & 1:1 &0.28 &&6&&0.922& 0.554&0\\    
        300 & 3 & 0 & 1:3 &0.28 &&3&&1.18&0.515&0.00210\\
        300 & 3 & 45 & 1:1 &0.28 &&-&&-&-&-\\
        300 & 3 & 45 & 1:3 &0.28 &&- && -&-&-\\
        300 & 5 & 45 & 1:1 &0.28 &&- && -&-&-\\
        300 & 5 & 45 & 1:3 &0.28 &&- && -&-&-\\
         300 & 10 & 45 & 1:1 &0.28 &&- && -&-&-\\
        500 & 1 & 0 & 1:1 &0.28 &&3&&     1.95&     0.378 & 0\\
         500 & 1 & 0 & 1:3  &0.28 &&1&&    1.26&     0.435 &0.00367\\
        500 & 1 & 45 & 1:1  &0.28 &&6&&   0.635&     0.526& 0\\
    500 & 1 & 45 & 1:3&0.28 &&3&&0.174&0.881&0.00079\\
         500 & 1 & 45 & 1:10  &0.28 &&-&& -&-&-\\
        500 & 3 & 0 & 1:3  &0.28 &&1&&    0.198&  0.556&     0\\
         500 & 3 & 45 & 1:3  &0.28 &&- && -&-&-\\
         500 & 5 & 45 & 1:3  &0.28 &&- && -&-&-\\
         700 & 1 &0 & 1:1  &0.28 &&1&&   0.517&     1.05&     0\\
        700 & 1 & 0 & 1:3  &0.28 &&2&&   1.19&     0.608&     0\\
        700 & 1 & 45 & 1:3  &0.28 &&4&&   0.170&     0.482&     0\\
        700 & 3 & 45 & 1:3  &0.28 &&- && -&-&-\\
         900 & 1 & 0 &1:1 &0.28 &&- && -&-&-\\
\hline
	\end{tabular}
	\label{tab:gadget-parameter}
	\tablecomments{The simulation configurations and the properties of the resulting DMDGs include: (1) relative collision velocity $v_{\rm coll}$  at a distance of 40 kpc, (2) pericentric distance of the collision $R_{\rm p}$, (3) relative angle between the two galactic disks, (4) initial mass ratio of two galaxies while the combined mass of the two is kept to $1.20\times10^{10}\,\rm M_{\odot}$, (5) initial gas fraction in each galaxy, $f_{\rm gas}=M_{\rm d, gas}/M_{200}$, (6) number of DMDGs identified at $t=870$ Myr (``-'' = none formed), (7) stellar mass of the most massive DMDG, (8) gas mass of the most massive DMDG, (9) dark matter mass of the most massive DMDG. See Section \ref{sec:3.3} for more information.	}
	\begin{tablenotes}
	\item[\textdagger]{\footnotesize This is the run shown in Figure \ref{fig:3}.}
	\item[\textdaggerdbl]{\footnotesize This is the run shown in the top row of Figure \ref{fig:5}.}
	\end{tablenotes}
	\end{threeparttable}
\end{table*}

\subsection{Idealized Fiducial Simulation: High-velocity Collision Between Two Dwarf Galaxies}\label{sec:3.2}

To investigate if a DMDG can be produced by a high-velocity collision of galaxies, we have tested an idealized binary collision of two gas-rich dwarf galaxies using both {\sc Enzo} and {\sc Gadget-2}. 
In our fiducial run in the present subsection, the galaxies (each of mass $M_{200} =5.95\times10^{9}{\,\rm M_{\odot}}$; see Section \ref{sec:2.1} for the detailed setup of a model galaxy) collide with a relative velocity of $300\,{\rm km\,s^{-1}}$ at 40 kpc distance.
We adopt a pericentric distance of 1 kpc (orbital eccentricity of 3.49) with both galactic disks lying on the plane of the collision orbit (i.e., coplanar collision).  
This combination of parameters depicts a near head-on collision in the two galaxies' first encounter. 
Figures \ref{fig:1} and \ref{fig:3} illustrate the results of the {\sc Enzo} and {\sc Gadget-2} simulations, respectively. 
Throughout the collision, bound structures such as galaxies are identified with the {\sc Hop} algorithm \citep[][applied only to the red star particles]{1998ApJ...498..137E}, and displayed as dashed circles of log$_{10}(M_{\star}/{\rm M_{\odot}})$ kpc radii.  
During and after the two galaxies make the first pericentric encounter at $t=0$ Myr, the collisionless dark matter halo (of blue particles in the bottom row) simply passes through one another.
Meanwhile, the warm gas in each galactic disk experiences highly supersonic collision. 
A significant portion of the gas particles lose their velocities and remain at the first contact point (the center of each panel) and along the tidally stripped gas stream (see the middle columns of Figures \ref{fig:1} and \ref{fig:3}). 
This process is reminiscent of the ``Bullet cluster'' where the collision of galaxy clusters decouples dark matter from dissipative baryons \citep[e.g.,][]{2006ApJ...648L.109C}. 

Along the gas stream dissociated from dark matter, the gas compressed by shock and tidal interaction cools and collapses to form star clumps in a timescale shorter than the local crossing time \citep{2004MNRAS.350..798B, 2009ApJ...706...67R}. 
The star clumps formed by Jeans instability in a long massive gas cylinder are in the shape of ``beads on a string''   \citep[middle columns of Figures \ref{fig:1} and \ref{fig:3};][]{1964ApJ...140.1529O, 1984pgs1.book.....F, 2010ASPC..423..177B}. Some of these star-forming clumps are large enough to eventually form galaxies with little dark matter. 
Some small clumps may merge with others by gravitational interaction and may form a larger structure (right columns of Figures \ref{fig:1} and \ref{fig:3}). 
Note that the results are for the most part reproducible between the two flavors of hydrodynamics codes, {\sc Enzo} and {\sc Gadget-2}.
At 870 Myr, the {\sc Enzo} run produced three DMDGs, while the {Gadget-2} run created two.  
Although the exact number or the locations of the collision-induced galaxies may be different between the two codes, their overall properties are similar, equally devoid of dark matter.  
The most massive ``collision-induced'' DMDG in the {\sc Enzo} run (Figures \ref{fig:1} and \ref{fig:2}) holds $1.60\times10^{8}\,{\rm M_{\odot}}$ in stars and  $4.17\times10^{7}\,{\rm M_{\odot}}$ in gas within 5 kpc radius, but only $2.71\times10^{5}\,{\rm M_{\odot}}$ in dark matter.  
In the {\sc Gadget-2} run (Figures \ref{fig:3} and \ref{fig:4}), the most massive DMDG acquired $2.16\times10^{8}\,{\rm M_{\odot}}$ in stars, $4.24\times10^{7}\,{\rm M_{\odot}}$ in gas, but merely $1.55\times10^{5}\,{\rm M_{\odot}}$ in dark matter.
The masses and compositions of these DMDGs are largely consistent with the estimated values for NGC1052-DF2 and NGC1052-DF4. 
However, these DMDG candidates have more gas than the reported gas mass of NGC1052-DF2,  $\lesssim 3.15\times10^{6}\,{\rm M_{\odot}}$ \citep{2019MNRAS.482L..99C}. 
One possible explanation is that most of gas in DMDGs can be lost after the collision in their violent environment. 
We do find that this phenomenon could happen in a collision event near a massive host (see Section \ref{sec:4.3}).
The most massive DMDGs tend to stay at the center of the image because the colliding gas --- which later forms stars --- has lost its incoming velocity and stalled at the contact surface.  

Shown in Figures \ref{fig:2} and \ref{fig:4} (left and  middle panels) are the properties of the most massive DMDG in the {\sc Enzo} and {\sc Gadget-2} run, respectively.  
The star formation rate (SFR;  left panels) is calculated for the newly-spawned star particles within 20 kpc from the most massive DMDG from $t$= -130 to 870 Myr.  
For both runs, the member star particles are born during and after the first encounter at $t=0$ Myr, implying that almost the entirety of the stellar mass in the DMDG is the direct result of collision-induced star formation.  
In particular, in the {\sc Enzo} simulation, the collision dramatically boosts the SFR to a maximum of $1.4\,\,{\rm M_{\odot}}{\rm yr}^{-1}$ when two galaxies are in a pericentric approach. 
For the {\sc Gadget-2} simulation, the SFR evolves more smoothly without a dramatic burst or drop.

Meanwhile, the middle panels of Figures \ref{fig:2} and \ref{fig:4} display the cumulative mass profile of the most massive DMDG at 870 Myr after the first pericentric encounter. 
For both {\sc Enzo} and {\sc Gadget-2} runs, the most massive DMDG is with stellar mass $\gtrsim 10^{8}\,{\rm M_{\odot}}$ but very little dark matter. 
It should be noted, however, that the internal structures of the DMDGs appear different between the codes.  
{\sc Enzo}'s DMDG (Figure \ref{fig:2}) has a more dispersed stellar distribution than {\sc Gadget-2}'s (Figure \ref{fig:4}) for  which $\sim 99$\% of the stellar mass is within 1 kpc from the center. 
The difference is likely due to our limited numerical resolution (80 pc; see Section \ref{sec:2.1}) --- sufficiently high to resolve the multiphase interstellar medium (ISM) and the ensuing collision-induced star formation, but not enough to resolve the internal dynamics of a dwarf-sized galaxy.  
The limited resolution tend to compound the discrepancy between the hydrodynamics solvers. Because of this reason, the detailed morphology of the resulting DMDGs such as the surface brightness and the number of globular clusters will be beyond the scope of the present paper.  By the same token, the simulated collision-induced DMDGs may not be straightforwardly compared with NGC1052-DF2 or NGC1052-DF4 for its morphology and/or internal structure  (see, however, our future work in Section \ref{sec:6.2}).  
Instead we will focus only on the overall properties of DMDGs observed in the simulations such as their mass and  composition.
Finally, the right panels of Figures \ref{fig:2} and \ref{fig:4} present the probability distribution functions (PDFs) of the gas density and temperature within 20 kpc from the center of the most massive DMDG at $t = 420$ Myr. 
Both {\sc Enzo} and {\sc Gadget-2} runs exhibit cold dense gas in the lower right corner above the star formation threshold density, $n_{\rm H, \, thres}$, which indicates continuous collision-induced star formation since the pericentric approach (as shown in the SFR, left panels of Figures \ref{fig:2} and \ref{fig:4}).  
A noticeable difference exists between the two runs, however,  such as the lesser warm gas in {\sc Gadget-2}, likely a result of limited resolution discussed in the preceding paragraph.  

\begin{figure*}
    \centering
    \includegraphics[width=0.95\textwidth]{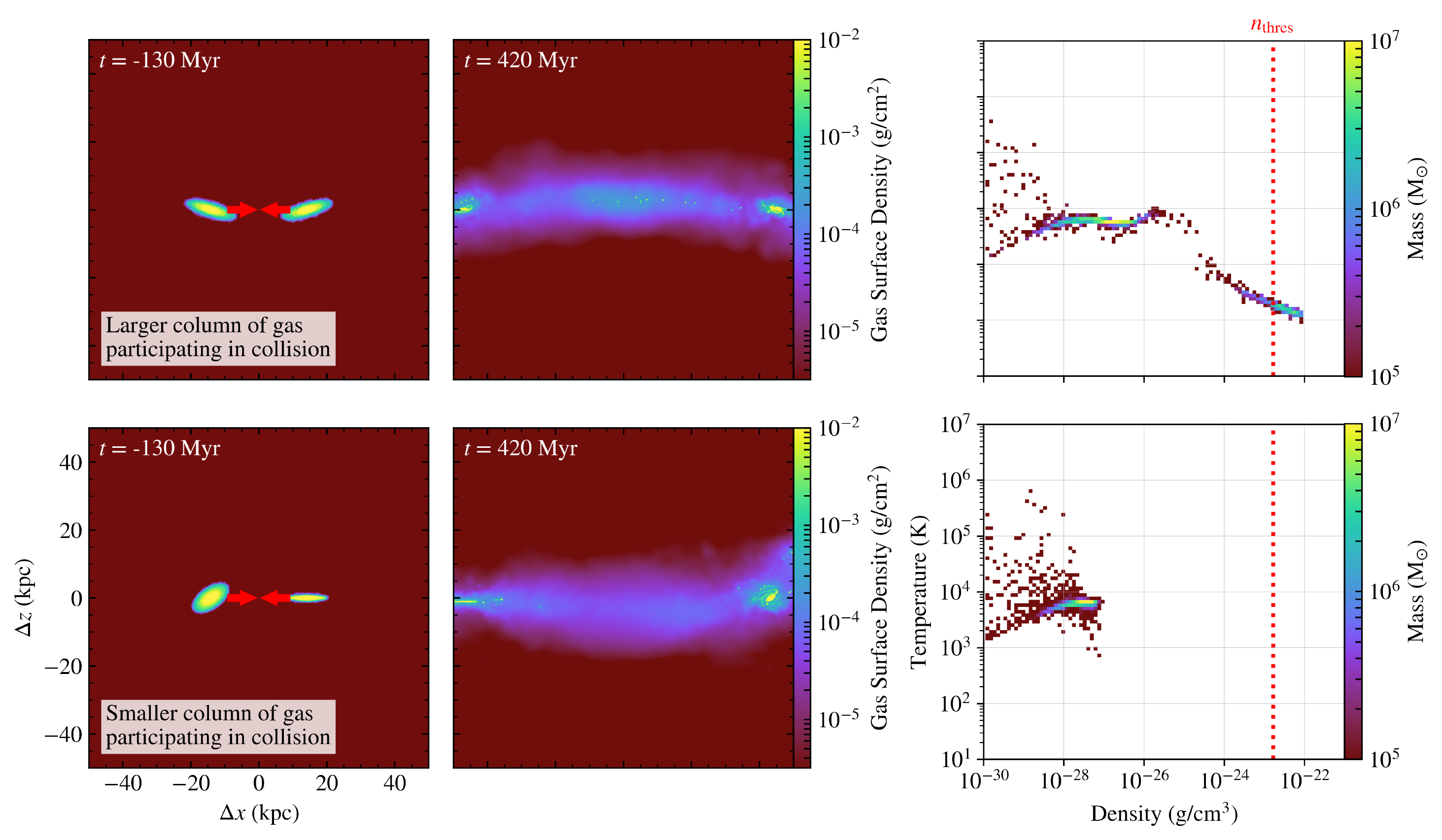}
    \vspace{-1mm}
    \caption{ 
    A set of two idealized collision simulations of two identical gas-rich dwarf galaxies with $M_{200} = 5.95\times10^{9}\,{\rm M_{\odot}}$ each, where the disks of the two galaxies make a 45$^{\circ}$ angle. 
    However, depending on the orientations of the disks with respect to the orbital plane, the amount of disk gas participating in the collision may differ.  
    {\it Top}: the tilts of the disks with respect to the orbital plane ($x$-$y$) are symmetric (22.5$^{\circ}$). 
    This is the configuration used in the runs with 45$^{\circ}$ in Table \ref{tab:gadget-parameter}.   
    {\it Bottom}: only one of the disks is tilted with respect to the orbital plane by 45$^{\circ}$. 
    In this configuration, only a small portion of the disk gas has a chance to directly clash into the other disk in the first near head-on collision.
    The run in the {\it top row} produces two DMDGs, while the run in the {\it bottom row} forms none, as shown in the edge-on gas surface densities  in the {\it left two columns} at $t = -130$ and 420 Myr.  
    In the {\it right column} are 2-dimensional density-temperature PDFs at $t=420$ Myr for the gas within 20$\,$kpc from the center of the most massive dark matter deficient object (a DMDG in the {\it top row}, a diffuse gas distribution in the {\it bottom row}). 
    The {\it red dashed line} denotes the star formation threshold density, $n_{\rm H, \, thres} = 10\, {\rm cm}^{-3}$. 
    See Section \ref{sec:3.3} for more information.
  }
    \vspace{3mm}
    \label{fig:5}
\end{figure*}

\subsection{Parameter Study: In What Type of Galaxy Collision Can A DMDG Form?}\label{sec:3.3}

We have shown that our fiducial high-velocity collision of two gas-rich galaxies can produce a DMDG with $M_{\star} \gtrsim 10^{8}\,{\rm M_{\odot}}$.  
We are thus naturally interested in exploring the parameter space for the collision to determine how often such DMDGs can form. 
Table \ref{tab:gadget-parameter} lists the suite of simulations we tested on {\sc Gadget-2} along with the properties of the resulting DMDGs.  
Parameters varied in the collision configuration include:  {\it (i)} relative collision velocity $v_{\rm coll}$ at a distance of 40 kpc (100, 300, 500, 700, 900 ${\rm km\,s^{-1}}$), {\it (ii)} pericentric distance $R_{\rm p}$ (1, 3, 5, 10 kpc), {\it (iii)} relative angle between the two galactic disks (0$^{\circ}$, 45$^{\circ}$),\footnote{The 45$^{\circ}$ angle configuration is illustrated in the top left panel of Figure \ref{fig:5}.} {\it (iv)} initial mass ratio of the two galaxies while the combined mass of the two is kept to $1.20\times10^{10}\,\rm M_{\odot}$ (1:1, 1:3, 1:10), {\it (v)} initial gas fraction in each galaxy, $f_{\rm gas} = M_{\rm d, gas}/M_{200}$ (0.02, 0.07, 0.13, 0.28).  

We find that one of the most crucial factors that determine the production of DMDGs is the amount of gas that actually participates in the collision and becomes {tidally and shock- compressed.} 
The amount of gas participating in the collision directly influences the amount of gas that loses its initial momentum and stalls at the first contact surface, thus determining the mass of the resulting collision-induced DMDG(s). 
It explains the following trends in Table \ref{tab:gadget-parameter}. 
\begin{itemize}
\item DMDGs are produced only when the initial gas fraction of the colliding galaxies is sufficiently large ($f_{\rm gas} > $  0.05).
The higher the initial gas fraction is, the more massive the resulting DMDGs tend to become.    
\item DMDGs are produced only when the pericentric distance is sufficiently small ($<$ 5 kpc for our experiments with $5.95\times10^{9}\,{\rm M_{\odot}}$ galaxies). 
A smaller pericentric distance means a larger ``cross-sectional area'' of the disk gas colliding into the other. 
\item Coplanar collision in which both galactic disks lie in the orbital plane ($x$-$y$; see Figure \ref{fig:3}) produces more DMDGs --- and more massive DMDGs --- than the one in which the disks make a 45$^{\circ}$ angle with respect to each other (see the top row of Figure \ref{fig:5}).  
It is because the coplanar collision gives a larger column of disk gas that a single gas particle will clash into.
\end{itemize}

The importance of gas mass that partakes in the collision in determining the formation of DMDGs is again demonstrated in an experiment shown in Figure \ref{fig:5}. 
Even when the two disks make the same 45$^{\circ}$ angle in both cases, the amount of disk gas participating in the collision may differ.
In the top row, two progenitor galaxies are initialized symmetrically, with the same 22.5$^{\circ}$ inclination angle with respect to the orbital plane (configuration used in the runs with 45$^{\circ}$ in Table \ref{tab:gadget-parameter}; case discussed in the last bullet point).  
In the bottom row, on the other hand, only one of the disks is tilted with respect to the orbital plane by 45$^{\circ}$.  
Therefore, the amount of disk gas actually participating in the collision is much larger in the top row.  
As a result, only the run in the top row produces multiple DMDGs, with its density-temperature PDF featuring a still significant amount of dense star-forming gas at $t=$ 420 Myr.    

Other factors also determine if a DMDG is produced or not:
\begin{itemize}
\item DMDGs are produced only when the relative collision velocity is in the [300, 700] ${\rm km\,s^{-1}}$ range.  
With the collision velocity of 100 ${\rm km\,s^{-1}}$, we find that the disk gas --- once decoupled after the first encounter --- is easily recaptured by the dark matter halos.  
The velocity of 900$\,{\rm km\,s^{-1}}$ brings the colliding gas to a state of excessively supersonic turbulence in which star formation is suppressed.
\item More DMDGs are produced in major mergers (mass ratio 1:1, 1:3) than in minor mergers (1:10).   
If one of the progenitor galaxies is significantly heavier than the other, the resulting collision-induced object tends to be much lighter than the heavier galaxy because of the small ``cross-sectional area'' of the disk gas clashing into the other (note that the combined mass of the galaxies is kept to $1.20\times10^{10}\,\rm M_{\odot}$). 
This means that the resulting DMDG, if any, could easily be recaptured by the heavier galaxy. 
\end{itemize}

\vspace{3mm}
\section{Looking For High-velocity Galaxy Collision Events in A Large Simulated Universe} \label{sec:4}

We have thus far focused on the idealized model of collision-induced DMDG formation using binary galaxy collision simulations  (Section \ref{sec:3}). 
But is this scenario really likely in our Universe? 
During the evolution of the Universe, how many high-velocity galaxy collisions have occurred in which a DMDG is expected to form? 
Armed with insights from our idealized experiments, we now attempt to answer the above questions by examining a large simulated universe  TNG100-1 introduced in Section \ref{sec:2.2}.

\subsection{Number of High-velocity Galaxy Collision Events in TNG100-1  That Could Have Produced DMDGs } \label{sec:4.1}

\begin{figure}
    \centering
    \includegraphics[width=0.48\textwidth]{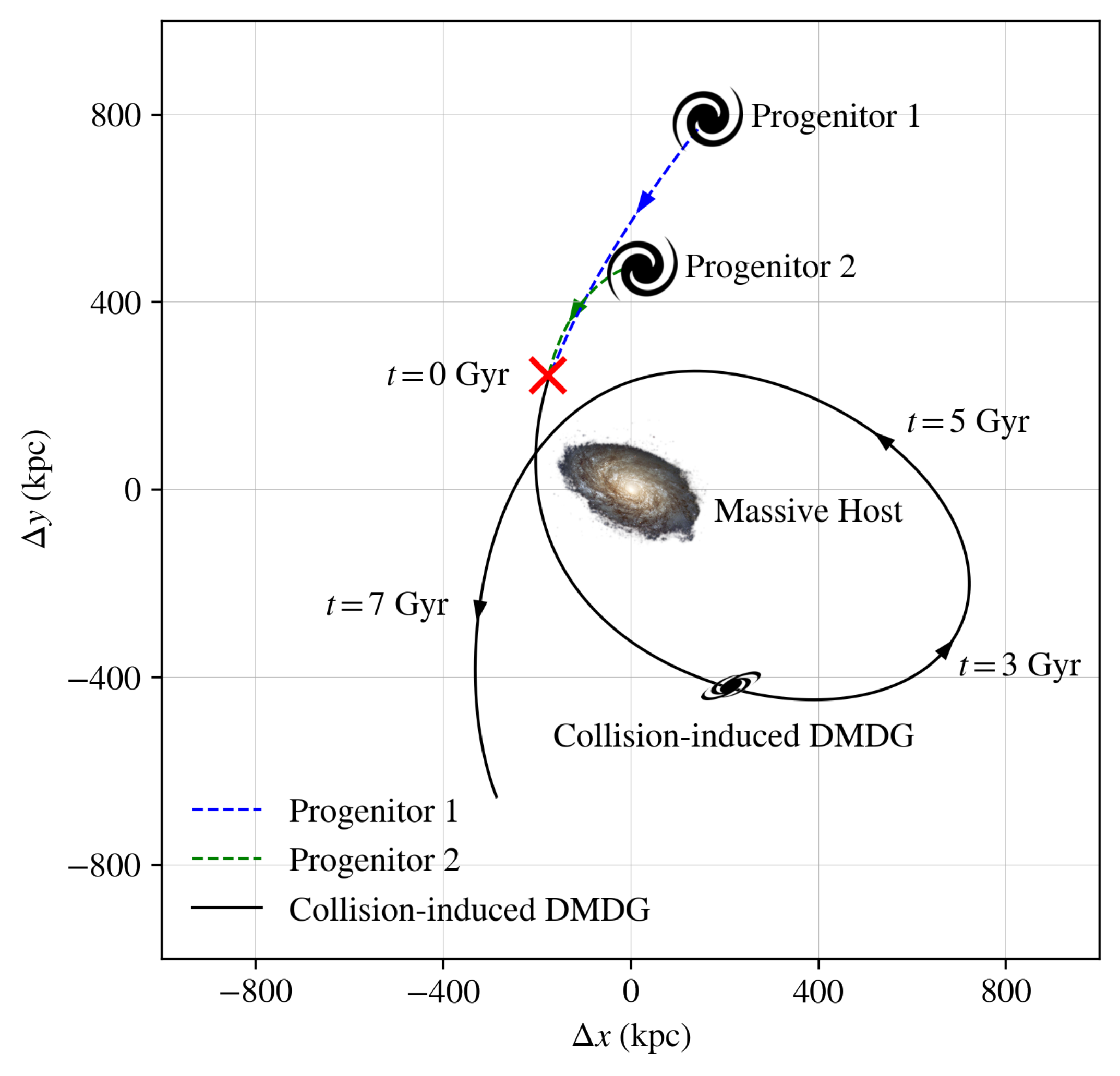}
    \caption{A predicted orbit of a ``possible'' collision-induced DMDG expected to form in one of the high-velocity galaxy collision events in  TNG100-1. 
    {\it Dashed lines}: the orbits of the two progenitor galaxies for 1 Gyr before the collision. 
    A massive host galaxy of $1.52\times10^{13}\,\rm M_{\odot}$ remains at the origin of this figure.  
    {\it Solid line}: the orbit of the collision-induced DMDG newly created at $t=0$ Gyr. 
    The DMDG is predicted to remain on a stable orbit around the massive host for the next 10 Gyr.
    See Section \ref{sec:4.2} for more information.
    }
    \vspace{10mm}
    \label{fig:6}
\end{figure}

In Section \ref{sec:3.3} and Table \ref{tab:gadget-parameter}, we explored the parameter space for a binary galaxy collision through a suite of idealized high-resolution simulations. 
Based on these insights, we now examine the {\sc SubFind} halo catalogues from TNG100-1 (see Section \ref{sec:2.2} for details) to count the number of galaxy collision events that {\it could} have produced collision-induced DMDGs.  
The search criteria include: 

\begin{itemize}
\item {\it (i)} The relative collision velocity between the two galaxies, when they are the closest in simulation snapshots, satisfies 200 $<v_{\rm coll}<$ 800$\,\,{\rm km\,s^{-1}}$, {\it (ii)} the pericentric distance $R_{\rm p}$ is less than 5 kpc, {\it (iii)} the initial mass ratio of the two galaxies satisfies 1/10 $<M_{1}/M_{2}<$ 10, and {\it (iv)} the initial gas fraction of the galaxies, $f_{\rm gas}$, is more than 0.05.
The conditions {\it (i)}-{\it (iv)} are based on the findings in our parameter study in Section \ref{sec:3.3}.\footnote{The disk inclination angle tested in Section \ref{sec:3.3} and Table \ref{tab:gadget-parameter} is not considered since disks cannot be distinctively defined for dwarf-sized galaxies in a relatively low-resolution simulation TNG100-1.} 
\item {\it (v)} The combined mass of the two galaxies, $M_{\rm total} = M_1 + M_2$, satisfies $3\times 10^{9}\,\rm M_{\odot}<$$M_{\rm total}$$<3\times10^{10} \,\rm M_{\odot}$. 
This condition is to approximately match the total mass chosen in the idealized runs ($1.20\times10^{10}\,\rm M_{\odot}$ in Sections \ref{sec:3.2}-\ref{sec:3.3}) that produces a DMDG similar to NGC1052-DF2 and NGC1052-DF4 (i.e., $M_{\star} \gtrsim 10^{8}\,{\rm M_{\odot}}$).  
\item {\it (vi)} The collision takes place between two satellite galaxies orbiting a more massive host galaxy  (see an example in Figure \ref{fig:6}). 
This condition is to mimic an environment where near head-on, high-velocity collision of galaxies might be more likely, and the resulting DMDG could be on a stable orbit around the massive host --- matching the fact that NGC1052-DF2 and NGC1052-DF4 are members of the NGC1052 group.  
The condition thus excludes the cases in which a collision occurs in a remote place far from any massive neighbors.\footnote{Among all the events that satisfy the conditions {\it (i)}-{\it (v)} in 95 snapshots from $z=10$ to 0.01, 181 events are excluded by the condition {\it (vi)}.} 
\item {\it (vii)} The satellite galaxies do not merge with the massive host right after the collision.  
This condition is to exclude the cases in which either of the two colliding satellites is dissolved in the host shortly after the collision, leaving little time for a DMDG to form and settle in a stable orbit.\footnote{Among all the events that satisfy the conditions {\it (i)}-{\it (v)} in 95 snapshots from $z=10$ to 0.01, 1485 events are excluded by the condition {\it (vii)}.} 
\end{itemize}

\begin{figure*}
    \centering
    \includegraphics[width=1.0\textwidth]{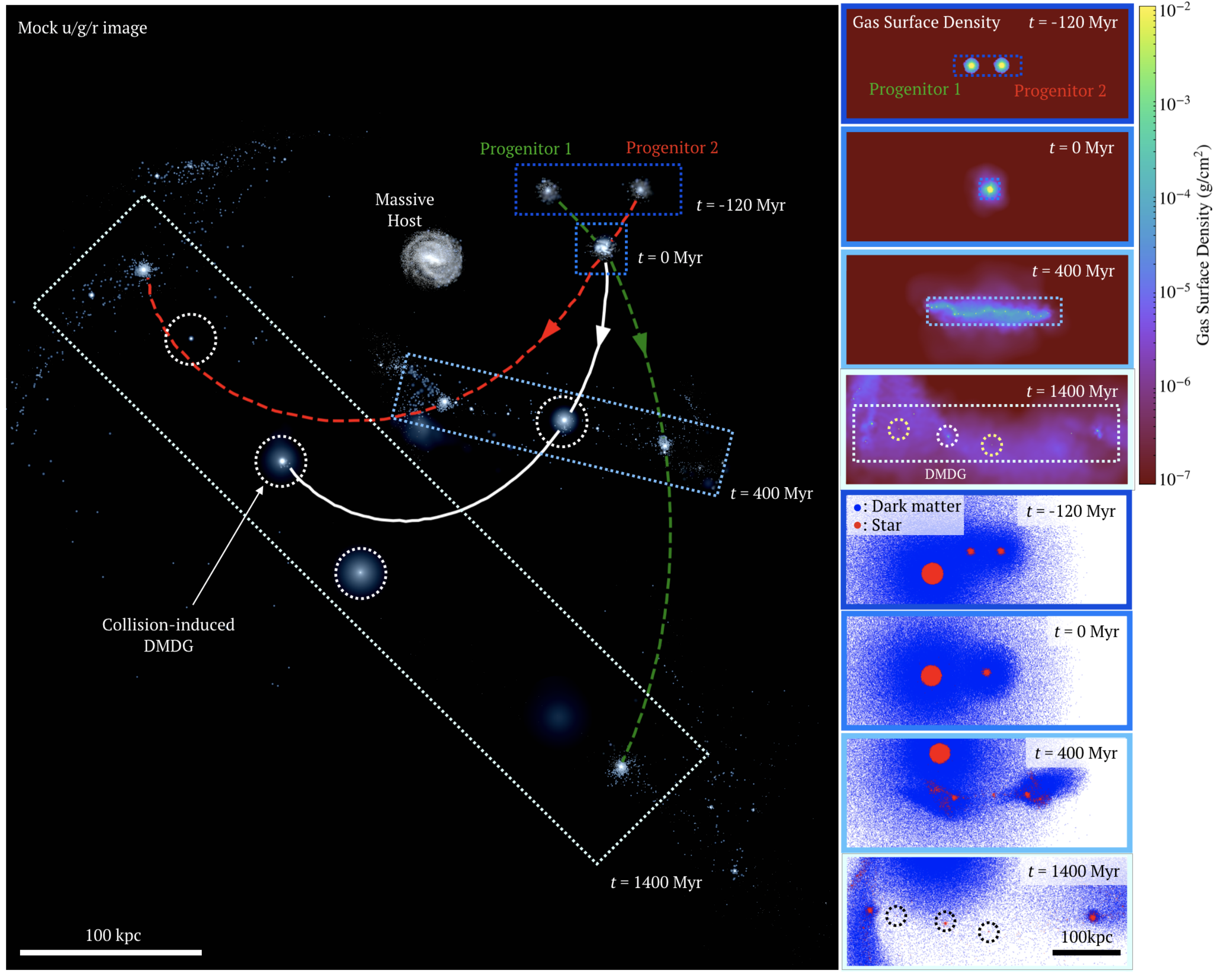}
    \caption{An idealized galaxy collision simulation on {\sc Gadget-2} demonstrating the ``collision-induced'' DMDG formation scenario.  
    The high-velocity ($\sim 300\,{\rm km\,s^{-1}}$) collision of two gas-rich, dwarf satellite galaxies near a massive host produces multiple DMDGs at $t=0$ Myr, the most massive one being with $M_{\star} = 6.03\times 10^{8}{\,\rm M_{\odot}}$.   
    {\it Left}:  mock {\it u/g/r} composites from the two colliding satellite galaxies and the DMDGs at four different epochs are superimposed on top of one another in a single frame.  
    The {\it white solid line} tracks the location of the most massive ``collision-induced'' DMDG, while the {\it green and red dashed lines} indicate the orbits of the two progenitor satellites.
    {\it Top right}: gas surface densities at the four epochs, $t=-120, 0, 400, 1400$ Myr.  
    {\it Bottom right}: dark matter particles in {\it blue} and newly-created star particles in {\it red} at the four epochs.   
    The DMDGs, marked with {\it dotted circles} at $t=1400$ Myr, are nearly devoid of dark matter ({\it blue} particles).  
    See Section \ref{sec:4.3} for more information about this figure and the experiment.}
    \label{fig:7}
    \vspace{7mm}
\end{figure*}

Applying these criteria to 95 halo catalogues from $z=10$ to 0.01, we obtain 248 galaxy collision events that {\it could} have produced collision-induced DMDGs.  
Whether the collision events actually produced DMDGs in the TNG100-1 run is a different question which will be addressed in Section \ref{sec:5}. 

\vspace{3mm}
\subsection{Example: How Would A DMDG-producing Galaxy Collision Event Look Like?} \label{sec:4.2}

In the previous subsection, we estimated the number of galaxy collision events in TNG100-1  in which a DMDG is expected to form.
This time we adopt even more stringent search criteria by including a condition about the orbit of the ``possible'' DMDG produced in the collision, and visualize one such event.  
Here we modify two existing criteria in Section \ref{sec:4.1}, and add one more:
\begin{itemize}
\item {\it (iii')} The initial mass ratio of two galaxies satisfies 1/4 $<M_{1}/M_{2}<$ 4, and {\it (iv')} the initial gas fraction of the galaxies is more than 0.2. 
\item {\it (viii)} The resulting DMDG attains a long-lasting stable orbit around the massive host, at least 50 kpc away from the center of the host at all times so that it does not suffer tidal disruption or stripping.  
For this condition, a crude prediction is made for the orbit of a ``possible'' DMDG (at this point we do not know the DMDG will really form) using the 4th-order Runge-Kutta method.\footnote{The following assumptions are made for this crude orbit prediction: {\it (a)} the system is fully described by the kinematics of the three bodies --- the two satellite progenitor galaxies and the massive host, {\it (b)} the motion of a satellite galaxy is determined only by the gravitational field of the host at the origin, {\it (c)} a DMDG forms at the center of mass of the two colliding galaxies (note that the DMDG forms when the gas from both galaxies collide perfectly inelastically), and {\it (d)} the newly-created DMDG tracks the center of mass of the two satellite galaxies afterwards.}     
\end{itemize}

With these stricter criteria, we discover three collision events in which a long-lasting DMDG is highly likely to form,  in the 95 TNG100-1 halo catalogues  from $z=10$ to 0.01. 
Figure \ref{fig:6} demonstrates the predicted orbit of a ``possible'' collision-induced DMDG in one of the three events. 
Two progenitor galaxies of masses $M_{\rm 200} = 1.84\times10^{10}\,\rm M_{\odot}$ and $3.41\times10^{10}\,\rm M_{\odot}$ each and the newly-created ``possible'' DMDG are all predicted to orbit the massive host of $M_{\rm 200}=1.52\times10^{13}\,\rm M_{\odot}$. 
The DMDG is expected to retain a stable orbit for more than 10 Gyr without approaching too close ($<$ 50 kpc) to the host galaxy.

\subsection{Idealized Simulation: High-velocity Collision Between Two Satellite Galaxies Near A Massive Host} \label{sec:4.3}

Did a galaxy collision event like Figure \ref{fig:6} really produce a long-lasting DMDG in TNG100-1?
Or, can a galaxy collision event like Figure \ref{fig:6} really produce a long-lasting DMDG in our Universe?
While we will search for an answer to the first question in Section \ref{sec:5}, here we will attempt to answer the second question by performing another idealized collision simulation. 
An initial condition similar to Figure \ref{fig:6} is used in both {\sc Enzo} and {\sc Gadget-2}.
A composite of simulation snapshots from the {\sc Gadget-2} run at four different epochs --- $t=-120, 0, 400, 1400$ Myr ---  is displayed in Figure \ref{fig:7}.  
Two dwarf satellite galaxies of $M_{\rm 200}=5.95\times10^{9}\,\rm M_{\odot}$ each stably orbiting a massive host of $M_{\rm 200}=1.07\times10^{12}\,\rm M_{\odot}$ are set on a collision course.\footnote{For the massive host at the center, the isolated disk galaxy initial condition from the {\it AGORA} Project  is adopted \citep[][see their Table 1]{2016ApJ...833..202K}, but without the gas disk for an expeditious calculation.}  
The initial configuration of this collision --- e.g., the positions and velocities of the satellites relative to the host --- is motivated by the three collision events found in Section \ref{sec:4.2} (e.g., Figure \ref{fig:6}). 
Progenitor 1 (2) is initially located 61.3 kpc (106.3 kpc) from the host with a velocity of 322 ${\rm km\,s^{-1}}$  (245 ${\rm km\,s^{-1}}$) relative to the host.  
The relative velocity between the two progenitors becomes 372 ${\rm km\,s^{-1}}$ at 50 kpc separation.
All other simulation setups such as star formation physics and spatial resolution are identical to those of the binary collision simulations in Sections \ref{sec:3.2}-\ref{sec:3.3}. 

Seen in the left panel of Figure \ref{fig:7} is the mock {\it u/g/r} composites from stellar and gas components at the four epochs, superimposed on top of one another.\footnote{Given the ages and metallicities of stars, the stellar spectra are computed using {\sc Starburst99}, then the attenuation of the flux due to the dust in the ISM is calculated assuming a constant dust-to-metal ratio and dust opacities for each band \citep[e.g.,][]{2005ApJ...625L..71H, 2014MNRAS.445..581H}.}
We find that our prescribed collision indeed produces multiple DMDGs with $M_{\star} > 10^{7}{\,\rm M_{\odot}}$ --- marked with dotted circles in the left panel --- via the same mechanism detailed in Section \ref{sec:3.2}.
The most massive DMDG (in a white dotted circle) has a stellar mass $M_{\star} = 6.03\times 10^{8}{\,\rm M_{\odot}}$.  
As can be seen in the sequence of four images along the orbit of the DMDGs in the right panels of Figure \ref{fig:7}, this most massive DMDG is almost completely devoid of dark matter at $t=1400$ Myr.  
The DMDG has gas mass $M_{gas} = 2.27\times 10^{7}{\,\rm M_{\odot}}$ at $t=1400$ Myr. 
However, we find that the tidal interaction with a massive host deprives the DMDG of most of its gas and leaves only $M_{gas} = 3\times 10^{6}{\,\rm M_{\odot}}$ by $t=3$ Gyr. This gas mass is consistent with the upper limit of gas in NGC 1052-DF2 reported by \cite{2019MNRAS.482L..99C}.
The DMDG is also stably bound  in the gravitational potential of the massive host for the entire simulation time of 10 Gyr, during which it is not tidally disrupted or distorted by the host.  
Its orbit has a radius of $\sim 100$ kpc (75 - 140 kpc) with the first sidereal period of 3.3 Gyr, which is largely consistent with our crude prediction (e.g., the solid line in Figure \ref{fig:6}). 
The {\sc Enzo} run also yields a qualitatively very similar result.  
We conclude that, in a numerical experiment with sufficiently high resolution (80 pc), a long-lasting DMDG can form near a massive host galaxy via a high-velocity collision of two gas-rich, dwarf satellite galaxies orbiting the same host.

\vspace{3mm}
\section{Looking for ``Collision-induced'' DMDGs in A Large Simulated Universe}\label{sec:5}

We have thus far demonstrated that in a numerical simulation with sufficiently high resolution (80 pc) a high-velocity collision of two gas-rich, dwarf galaxies can produce one or more DMDGs.  
It is also found that there is a significant number of galaxy collision events in the TNG100-1 universe in which a DMDG is expected to form.  
Now we turn our attention to whether the ``collision-induced'' DMDGs have really formed in the TNG100-1 run.   
While there have been attempts by several groups to examine if DMDGs are found in large simulated universes such as {\sc Illustris} and {\sc Eagle} \citep[e.g.,][]{2018arXiv180905938Y, 2019MNRAS.488.3298J,  2019MNRAS.489.2634H, 2019A&A...626A..47H}, they have mostly focused on the tidal stripping or the old TDG scenarios (see Section \ref{sec:1} for various DMDG formation scenarios proposed).\footnote{\cite{2018arXiv180905938Y} showed that $\sim$ 0.7\% of galaxies  with $M_{\star} \sim 2 \times 10^8\,{\,\rm M_{\odot}}$ (stellar mass similar to NGC1052-DF2's) in the {\sc Illustris-1} simulation \citep{2014MNRAS.445..175G} shows $M_{\star} > M_{\rm DM}$.  \cite{2019MNRAS.488.3298J} argued that 1-2\% of satellite galaxies with $10^9 {\,\rm M_{\odot}} < M_{\star}  <10^{10} {\,\rm M_{\odot}}$ are significantly dark matter deficient in the {\sc Illustris-1} and {\sc Eagle} simulations \citep{2015MNRAS.446..521S}. They also suggested that the majority of the DMDGs in these simulations might have suffered tidal stripping. \cite{2019A&A...626A..47H} surveyed the {\sc Illustris-1} volume to find that several DMDGs are TDGs. \label{footnote:previous-work}}   
To the best of our knowledge the ``collision-induced'' DMDG formation scenario has never been examined in any large-scale  simulation.  
In this section, we search for the evidence for this particular scenario in the TNG100-1 universe with two approaches --- first using its halo catalogues (Section \ref{sec:5.1}), then using its simulation snapshots themselves (Section \ref{sec:5.2}).

\begin{figure}
    \centering
    \includegraphics[width=0.46\textwidth]{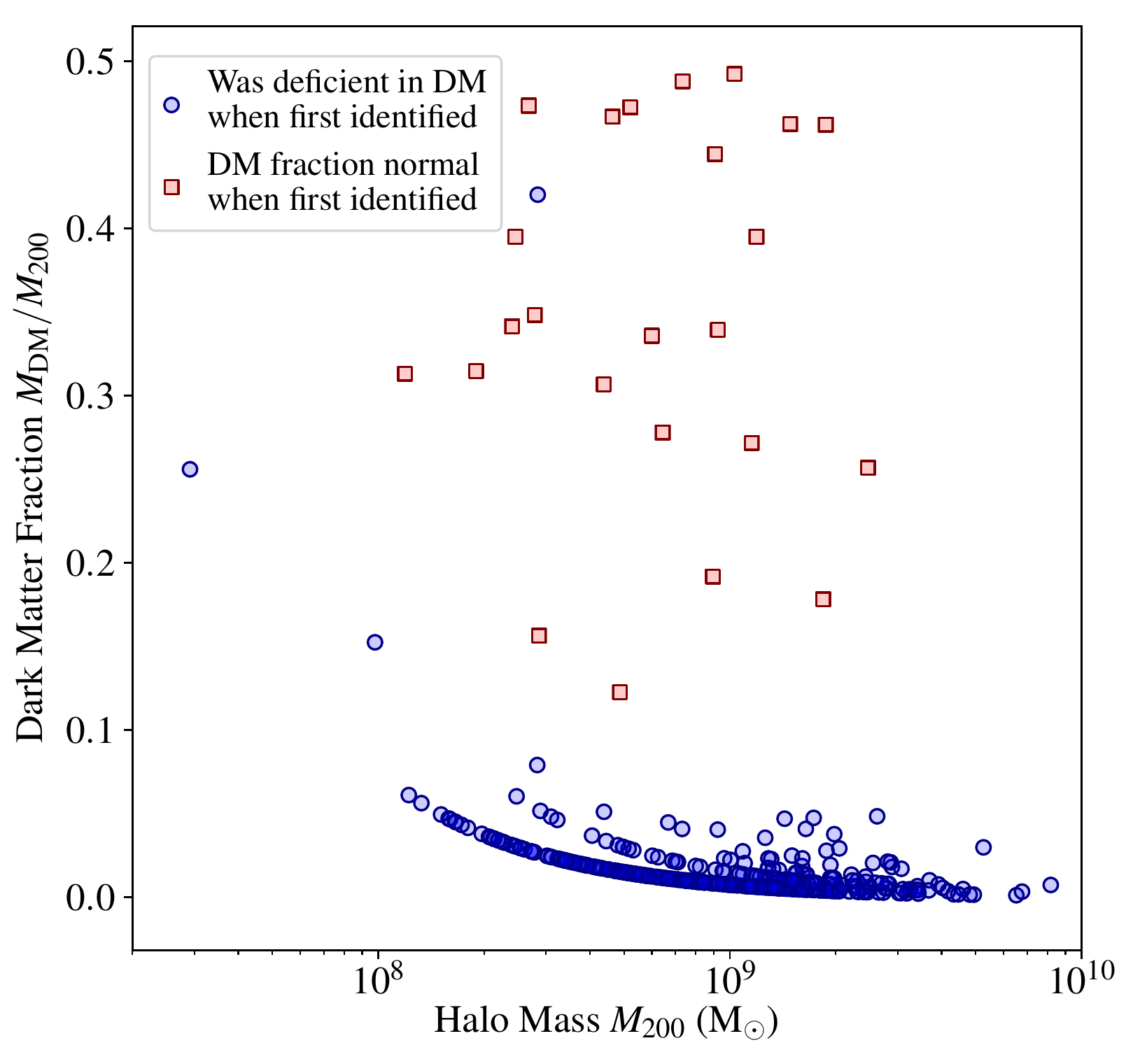}
    \vspace{-1mm}
    \caption{Dark matter fraction as a function of total mass for the dark matter deficient objects selected with the criteria detailed in Section \ref{sec:5.1} in the  TNG100-1 halo catalogue at $z=0$.  
    Of 435 such objects, 411 in {\it blue circles} were already dark matter deficient (i.e., $M_{\rm DM}/M_{200} < $  0.5) when they were first identified as subhalos at $z>0$.  
    Meanwhile, the remaining 24 objects in {\it red squares} had rather normal dark matter fractions (i.e., $M_{\rm DM}/M_{200} > $  0.5) when first identified, but have since lost dark matter as they evolve in time to $z=0$.  
    See Section \ref{sec:5.1} for more information.}
    \label{fig:8}
    \vspace{5mm}
\end{figure}

\subsection{Looking for Collision-induced DMDGs in the TNG100-1 Halo Catalogues}\label{sec:5.1}

In this subsection, we look for ``collision-induced'' DMDGs using the {\sc SubFind} halo catalogues and the {\sc SubLink} merger trees from the $z=0$ TNG100-1 dataset (Section \ref{sec:2.2}), but not the simulation snapshots themselves.  
We start by {\it (1)} identifying dark matter deficient objects that are of our interest --- DMDGs or otherwise ---  in the halo catalogues, then move to {\it (2)} examining the merger trees of their neighboring galaxies to find any evidence of collision-induced DMDG formation.  

First, the entire TNG100-1 halo catalogue at $z=0$ is surveyed to search for any satellite galaxies (subhalos) that are deficient in dark matter.  
To do so, very restrictive selection criteria are adopted as follows:
\begin{itemize}
\item {\it (i)} The galaxy's dark matter mass is less than 50\% of its total mass, and {\it (ii)} its stellar mass (or its gas mass) is more than $10^7\,\rm M_{\odot}$.  
The conditions {\it (i)}-{\it (ii)} are to match the first two items in our definition of DMDGs in Section \ref{sec:3.1}.  
\item {\it (iii)} The galaxy is located at $[10, 100] \,R_{\rm h,\, host}$ from the host galaxy, where  $R_{\rm h,\, host}$ is the stellar half mass radius of the host.  
Here, the host galaxy is defined as a massive halo with $M_{\star} > 10^9\,\rm M_{\odot}$ nearest to our target object.  
This condition is to exclude misidentified subhalos that are not galaxies, such as globular clusters \citep{Ploeckinger2018, 2019A&A...626A..47H}.
\item {\it (iv)} The galaxy has at least one dark matter particle.  
This condition is simply to ensure that our galaxies are included in the {\sc SubLink} merger trees.\footnote{Among the 4171 galaxies that satisfy the conditions {\it (i)}-{\it (iii)} at $z=0$, 3776 are excluded by the condition {\it (iv)}.} 
\end{itemize}

Applying these criteria to the $z=0$ halo catalogue, we identify 435 candidates for DMDGs.  
We categorize them into two groups, and consider how a ``collision-induced'' DMDG may manifest itself in each group: 

\begin{itemize}
\item The 411 objects in the first group (blue circles in Figure \ref{fig:8}) are dark matter deficient not only at $z=0$ but also when they were first identified as subhalos at $z>0$.  
In other words, these galaxies have been deficient in dark matter from the beginning of the main branch in its merger tree down to $z=0$.  
Most of the these galaxies have only one or two dark matter particles at $z=0$ as can be seen in Figure \ref{fig:8}. 
Upon further inspection, however, 213 of these {\it inherently} dark matter deficient objects are identified by {\sc SubFind} for the first time at $z=0$ --- either because they are of non-cosmological origin (i.e., not galaxy, but globular clusters or large gas clumps only temporarily misidentified as subhalos, according to our visual inspection), or because they cannot be reliably traced back in time  by {\sc SubLink} because there are too few particles.  
Indeed, only 56 objects out of 411 are flagged as galaxies.\footnote{${\tt SubhaloFlag} = 1$, an extra flag in the {\sc IllustrisTNG} halo catalogue, described in \citealt{2019ComAC...6....2N}.}
In the collision-induced DMDG formation, the DMDGs  {\it could} have formed when two more massive galaxies collided in their neighborhood (but did not merge).
We therefore search for the cases in which two other halos are located within 50 kpc from a newly-identified DMDG.\footnote{50 kpc is chosen based on the relative velocity of the colliding galaxies we consider in Sections \ref{sec:3} and \ref{sec:4} ($\sim 300 \,{\rm km\,s^{-1}}$), and the temporal resolution between the TNG100-1 snapshots ($\gtrsim 0.2$ Gyr).}   
Only 19 such cases are found, and none of them shows a sign of a recent high-velocity interaction between the two nearby galaxies.\footnote{We have also tested the hypothesis with the baryon-based merger trees {\sc SubLink-Gal} that can trace (especially) the dark matter deficient objects further back in time using baryonic particles. Still, none of the objects at $z=0$ is found to have formed via a galaxy collision.}$^{,}$\footnote{Upon tracking the particles of 19 inherently dark matter deficient objects, we found that these are not TDGs, but substructures of a massive galaxy. A few subhalos are related to the tidal interaction with other massive galaxies but they do not seem to be TDGs.}
\item The remaining 24 galaxies in the second group (red squares in Figure \ref{fig:8}) were {\it not} dark matter deficient  when they were first identified as subhalos at $z>0$, but have since lost dark matter as they evolve to $z=0$.  
This group of galaxies is what other authors have partially studied.$^{\ref{footnote:previous-work},}$\footnote{\cite{2020MNRAS.491.1278S} argued that in the particle-based simulations like {\sc Illustris}, {\sc IllustrisTNG} and {\sc Eagle}, a numerical error may strip a few subhalos of their dark matter at the edge of the simulation box.}
In a collision-induced DMDG formation scenario, on the other hand, a DMDG's drastic decrease in dark matter ($M_{\rm DM}/M_{200} $ moving below 0.5) {\it could} be associated with the moment of a major merger between two neighboring galaxies.  
We, however, find that for none of the 24 galaxies, a nearby major merger (mass ratio $>$ 1:5) --- if exists at all --- coincides with the decrease in its dark matter fraction.   
\end{itemize}

Hence, we can conclude that the few DMDGs found in the $z=0$ halo catalogue are not related to the high-velocity collision of nearby galaxies.  
Our conclusion is puzzling because we know that high-velocity galaxy collisions exist in the TNG100-1 volume  (Section \ref{sec:4}) and that such collisions could produce DMDGs (Section \ref{sec:3}). 
We now turn to the second method to search for the collision-induced DMDGs, and seek to acquire insights into what really happened in the  TNG100-1 run.

\begin{figure*}
    \centering
    \vspace{-2mm}
    \includegraphics[width=0.9\textwidth]{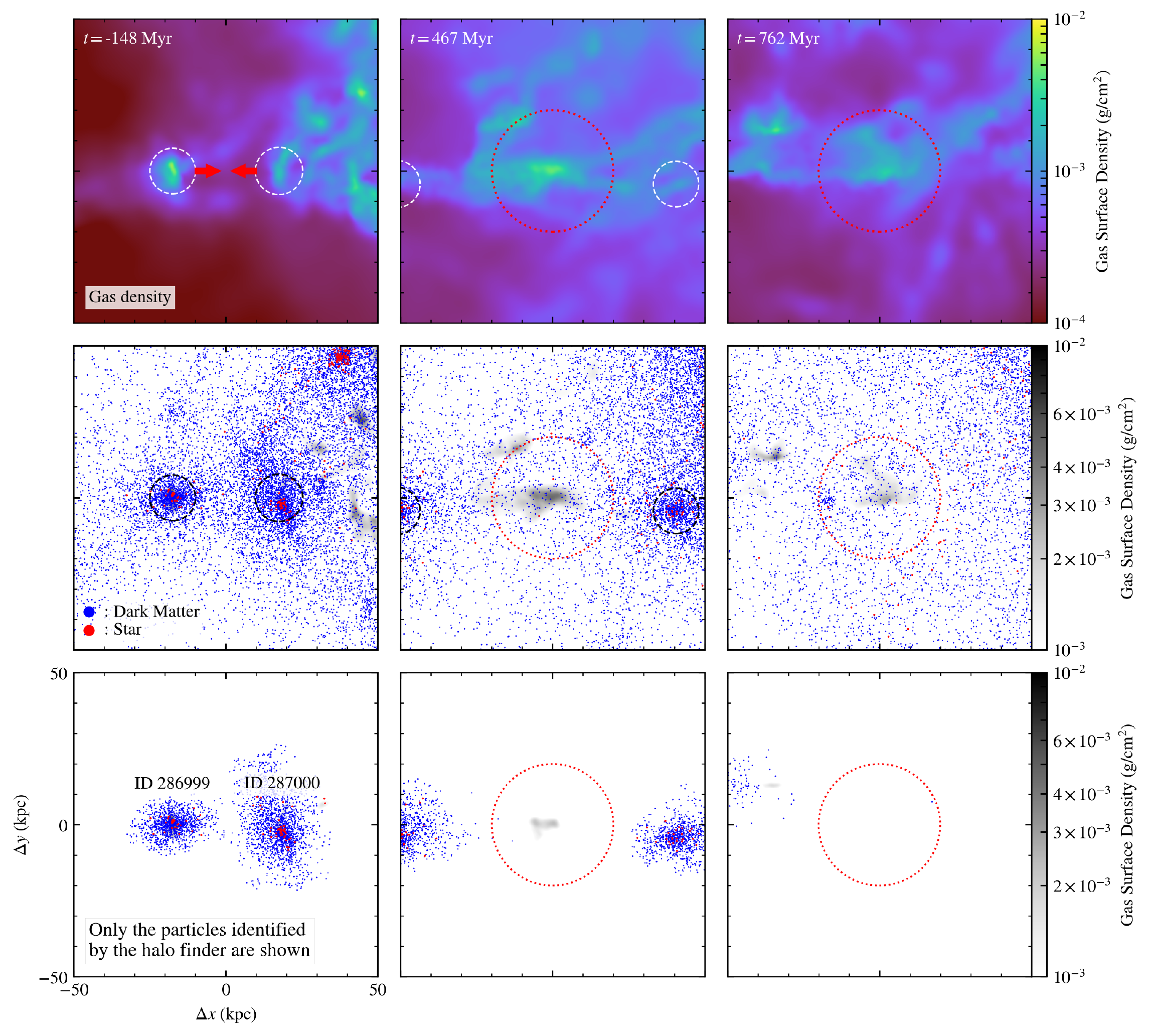}
    \vspace{-2mm}
    \caption{Similar to Figures \ref{fig:1} and \ref{fig:3}, but depicting one of the high-velocity galaxy collisions found in the TNG100-1 run at three epochs, where $t$ is the time since the two galaxies' pericentric approach.
    The two galaxies have masses $M_{200}=9.48\times10^{9}\,\rm M_{\odot}$ ($f_{\rm gas} \sim 0.06$; moving from left to right) and $M_{200}=1.47\times10^{10}\,\rm M_{\odot}$ ($f_{\rm gas} \sim 0.07$; from right to left), approaching at $227\,{\rm km\,s^{-1}}$ when 36 kpc apart. 
    The estimated pericentric distance is 6.2 kpc.  
    {\it Top row}: gas surface density. 
    {\it Middle row}: dark matter particles in {\it blue} and newly-created star particles {\it red}.
    Galaxies identified by the {\sc SubFind} algorithm are marked with {\it black or white dashed circles} of log$_{10}(M_{\star}/{\,\rm M_{\odot}})$ kpc radii.      
    At $t=0$ Myr, a dark matter deficient object --- not a DMDG, but a giant gas clump --- forms as a result of the collision (marked with a {\it red dotted circle} of 20 kpc radius, $M_{\rm gas,\,<5kpc} = 4.75\times 10^{8}{\,\rm M_{\odot}}$; see Figure \ref{fig:10}).
    {\it Bottom row}: only the particles that belong to the {\sc SubFind} halos are shown, in which the collision-induced gas clump in the {\it middle row} is not seen. The small gas clump in the {\it middle panel} of the {\it bottom row} is not identified as an independent halo but as a part of one of the two neighboring halos.
See Section \ref{sec:5.2} for more information.}
    \label{fig:9}
    \vspace{-2mm}
\end{figure*}
\begin{figure*}
    \centering
    \includegraphics[width=0.79\textwidth]{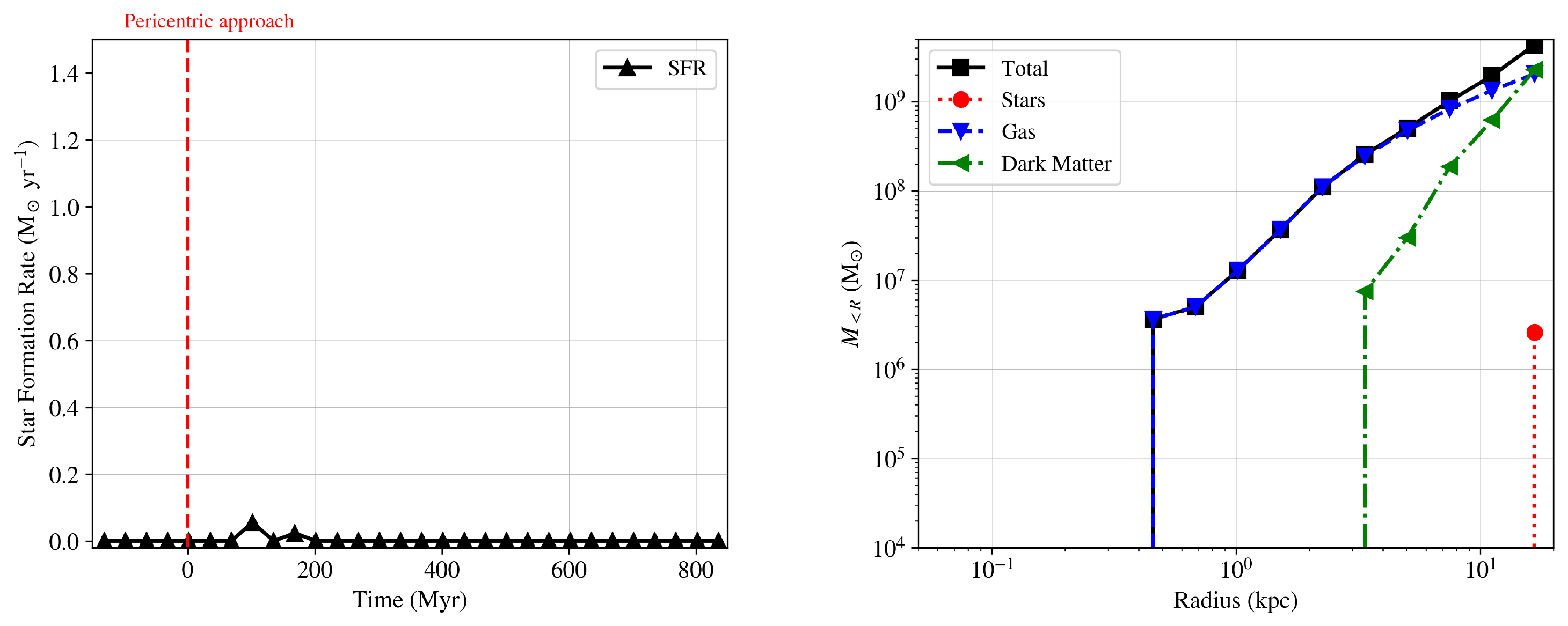}
    \vspace{-1mm}    
    \caption{Similar to Figures \ref{fig:2} and \ref{fig:4}, but for the dark matter deficient gas clump in Figure \ref{fig:9}.
    {\it Left}: the SFR within 20 kpc from the clump.   
    {\it Right}: the cumulative mass profile at $t=762$ Myr, showing little dark matter or stars within 5 kpc.  
    The $y$-axes are kept identical for Figures \ref{fig:2},  \ref{fig:4}, \ref{fig:10} and \ref{fig:12} for easier comparison.
    }
    \label{fig:10}
    \vspace{-2mm}
\end{figure*}

\begin{figure*}
    \centering
    \includegraphics[width=0.95\textwidth]{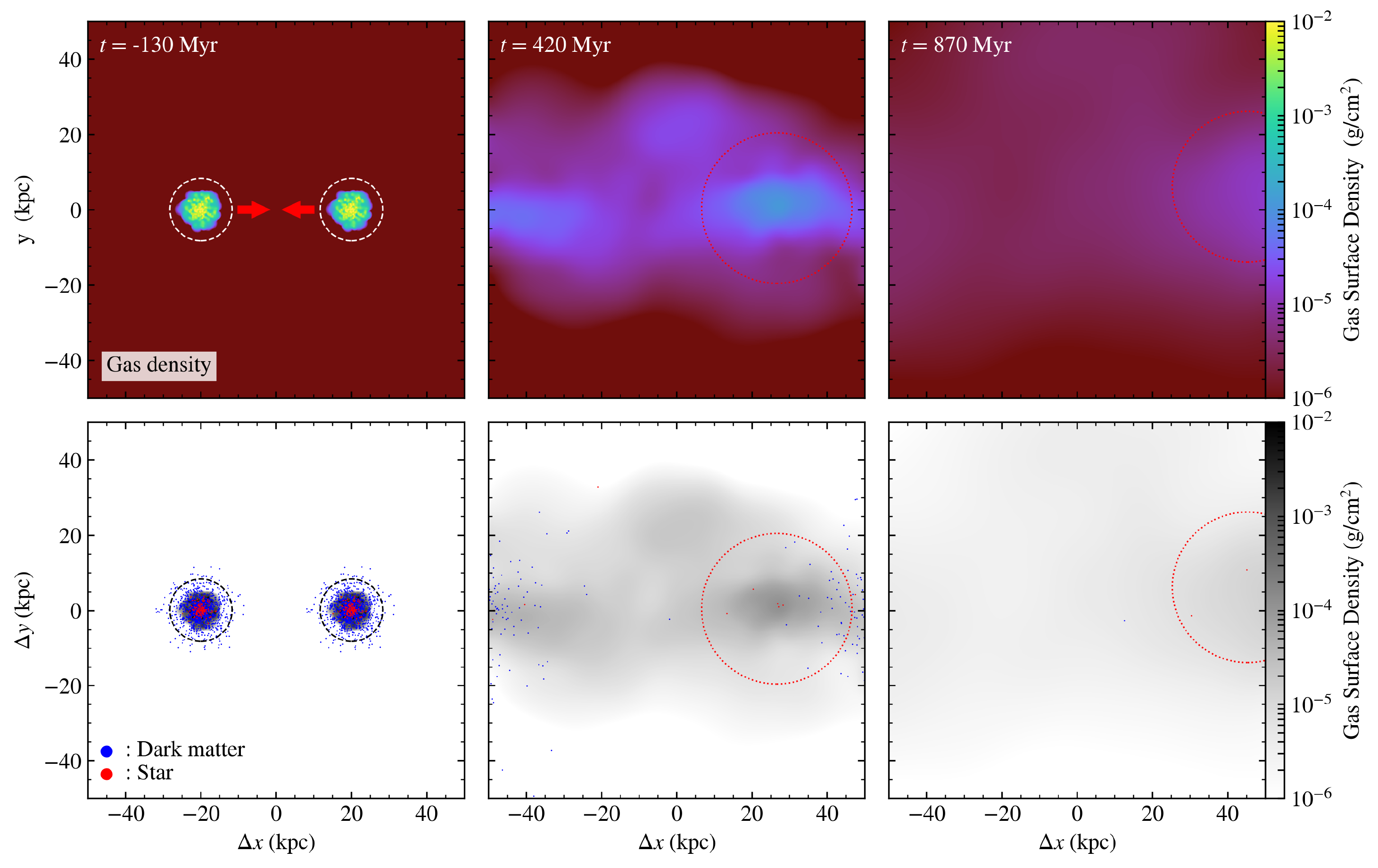}
    \vspace{-1mm}    
    \caption{
    Similar to Figures \ref{fig:1}, \ref{fig:3} and \ref{fig:9}, but for an idealized galaxy collision simulation performed on {\sc Gadget-2} with lowered spatial resolution ($\epsilon_{\rm grav}= 185$ pc as opposed to 80 pc in Figure \ref{fig:3}) to match TNG100-1 (Figure \ref{fig:9}). 
    Total gas mass of the two galaxies is set to be approximately equal to that of Figure \ref{fig:9} for comparison.  
    All other collision parameters are identical to those in its higher-resolution counterpart in Figure \ref{fig:3} (see also Table \ref{tab:gadget-parameter-2}).
    At $t=0$ Myr, the collision produces a dark matter deficient object --- not a DMDG, but a gas clump (marked with a {\it red dotted circle} of 20 kpc radius) --- which then quickly loses its mass by $t=870$ Myr ($M_{\rm gas,\,<20kpc} = 4.48\times 10^{7}{\,\rm M_{\odot}}$; see Figure \ref{fig:12}).
   See Section \ref{sec:5.3} for more information.
    }
    \label{fig:11}
    \vspace{-1mm}    
\end{figure*}

\begin{figure*}
    \centering
    \vspace{-1mm}    
    \includegraphics[width=0.79\textwidth]{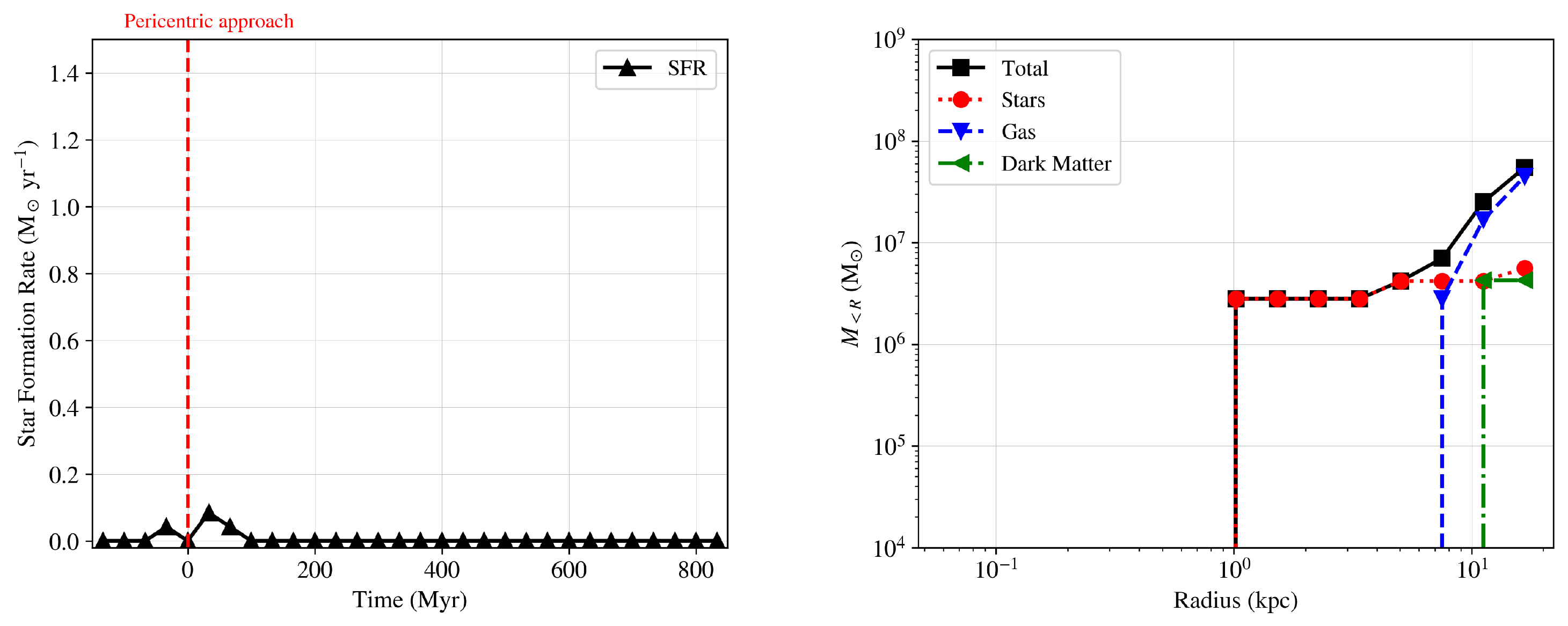}
    \vspace{-1mm}        
    \caption{
    Similar to Figures \ref{fig:2}, \ref{fig:4} and \ref{fig:10}, but for the dark matter deficient gas clump in the low-resolution {\sc Gadget-2} run shown in Figure \ref{fig:11}.  
    As in Figure \ref{fig:10}, the gas clump shows very little star formation activity ({\it left}), in clear contrast with the higher-resolution simulations in Figures  \ref{fig:2} and \ref{fig:4}.  
    The clump disintegrates and loses its mass by $t=870$ Myr ({\it right}).
    The $y$-axis ranges of these panels are kept identical for Figures \ref{fig:2},  \ref{fig:4}, \ref{fig:10} and \ref{fig:12} for easier comparison.
    }
    \label{fig:12}
    \vspace{-1mm}
\end{figure*}

\begin{table*}
\centering
\vspace{-3mm}
\caption{A suite of idealized galaxy collision simulations on {\sc Gadget-2} to test the effect of numerical resolution and star formation physics for the formation of collision-induced DMDGs.}

\begin{threeparttable}
\vspace{1mm}
\begin{tabular}{@{}cccccccccccc@{}}
\toprule
\multicolumn{5}{c}{Collision parameters}&&\multicolumn{2}{c}{Resolution-related } && Most massive \\
\multicolumn{5}{c}{}&&\multicolumn{2}{c}{parameters} && DMDG \\
\cline{1-5}\cline{7-8}\cline{10-10}
Collision velocity &  Pericentric distance & Disk angle & Mass ratio & Gas fraction &&  Resolution & Star formation threshold && $M_{\star}$ \\
 (${\rm km\,s^{-1}}$)&(kpc)&($^{\circ}$)& & &&$\epsilon_{\rm grav}$ (pc) &  $n_{\rm H,\,thres}$ (cm$^{-3}$)  && ($10^8\,\rm M_{\odot}$)  \\
 (1)&(2)&(3)&(4)&(5)&&(6)&(7)&&(8)\\
\hline
300& 1& 0& 1:1& 0.28&& 80& 10&& 2.16\tnote{\textcolor{red}{*, \textdagger}} \\
300& 1& 0& 1:1&0.13&& 80& 10&&0.993\tnote{\textcolor{red}{*}}\\
300& 1& 0& 1:1&0.07&& 80& 10&& 0.303\tnote{\textcolor{red}{*}}\\
300& 1& 0& 1:1&0.02&& 80& 10&& -\tnote{\textcolor{red}{*}}\\
300& 1& 0& 1:1&0.42&& 185& 0.13&& 4.90\\
300& 1& 0& 1:1&0.35&& 185 & 0.13&& 3.68\\
300& 1& 0& 1:1&0.28&& 185& 0.13&& 0.746\\
300& 1& 0& 1:1&0.23&& 185& 0.13&& -\\
300& 1& 0& 1:1&0.14&& 185& 0.13&& -\tnote{\textcolor{red}{\textdaggerdbl}}\\
300& 1& 0& 1:1&0.42&& 370& 0.13&& 0.626\\
300& 1& 0& 1:1&0.35&& 370& 0.13&& 464\\
300& 1& 0& 1:1&0.28&& 370& 0.13&& -\\
300& 1& 0& 1:1&0.24&& 370& 0.13&& -\\  
300& 1& 0& 1:1&0.47&& 740& 0.13&& 0.591\\
300& 1& 0& 1:1&0.42&& 740& 0.13&& 0.158\\
300& 1& 0& 1:1&0.35&& 740& 0.13&& -\\
300& 1& 0& 1:1&0.28&& 740& 0.13&& -\\
\hline

	\end{tabular}
	\label{tab:gadget-parameter-2}
	\vspace{1mm}
	\tablecomments{The simulation configurations and the properties of the resulting DMDGs include: (1) relative collision velocity $v_{\rm coll}$  at a distance of 40 kpc, (2) pericentric distance of the collision $R_{\rm p}$, (3) relative angle between the two galactic disks, (4) initial mass ratio of two galaxies while the combined mass of the two is kept to $1.20\times10^{10}\,\rm M_{\odot}$, (5) initial gas fraction in each galaxy, $f_{\rm gas} = M_{\rm d, gas}/M_{200}$, (6) the Plummer equivalent gravitational softening length $\epsilon_{\rm grav}$, (7) star formation threshold density $n_{\rm H,\,thres}$, (8) stellar mass of the most massive DMDG  (``-'' = none formed). 
	See Section \ref{sec:5.3} for more information.}
	\vspace{1mm}
	\begin{tablenotes}
	\item[*]{\footnotesize These are the same runs from Table \ref{tab:gadget-parameter}.}
	\item[\textdagger]{\footnotesize This is the run shown in Figure \ref{fig:3}.}
	\item[\textdaggerdbl]{\footnotesize This is the run shown in Figure \ref{fig:11}.}
	\end{tablenotes}
	\end{threeparttable}
	\vspace{1mm}
\end{table*}

\subsection{Looking for Collision-induced DMDGs in the TNG100-1 Simulation Snapshots}\label{sec:5.2}

The second approach in this subsection to look for any hint of ``collision-induced'' DMDGs is to utilize the TNG100-1 snapshots themselves.  
Due to its small size and unconventional dynamical properties, a newly-formed DMDG in between the two colliding galaxies might not have been identified by the {\sc SubFind} halo finder.  
Or worse, the collision may not have formed a DMDG in the first place for numerical reasons.  
To test these hypotheses, we explore the ``subboxes'' of TNG100-1 with finer output intervals (a few Myr as opposed to $\gtrsim 0.2$ Gyr for the original TNG100-1 snapshots; see Section \ref{sec:2.2}) that is crucial to trace the kinematics of gas in colliding galaxies, and identify the ``possible'' formation sites of DMDGs.

To locate high-velocity galaxy collision events in thousands of timesteps in the ``subbox'', we apply the search criteria that are very similar to those in Section \ref{sec:4.1}.\footnote{The search criteria have to be slightly relaxed from Section \ref{sec:4.1} to have at least a few candidate events to emerge.  Note that the volume of a ``subbox'' is only 0.11\% of the TNG100-1 volume.  Specifically, the condition {\it (ii)} in Section \ref{sec:4.1} is relaxed to: {\it (ii')} the pericentric distance $R_{\rm p} < $ 10 kpc.}
After visually inspecting each of the eight candidates in the two subboxes available, we identify one collision event in the  {\sc TNG100-1-Subbox0} volume that produced {\it any} dark matter deficient object --- in this case, a giant gas clump. 
The left column of Figure \ref{fig:9} illustrates the two galaxies in this event at $z\sim 1.3$. 
Each galaxy has $M_{200}=9.48\times10^{9}\,\rm M_{\odot}$ with $f_{\rm gas} \sim 0.06$ (or  $M_{\rm gas}=6.60\times10^{8}\,\rm M_{\odot}$; moving from left to right) and $M_{200}=1.47\times10^{10}\,\rm M_{\odot}$ with $f_{\rm gas} \sim 0.07$ (or $M_{\rm gas}=1.02\times10^{9}\,\rm M_{\odot}$; moving from right to left). 
The two galaxies approach at a relative velocity of $227\,{\rm km\,s^{-1}}$ at 36 kpc distance, and with an estimated pericenter distance $6.2\,{\rm kpc}$, largely compatible with idealized simulations  in Sections \ref{sec:3.2}-\ref{sec:3.3}, and \ref{sec:4.3}. 

During the high-velocity collision, dark matter halos pass through each other while a significant portion of gas experiences ram pressure and remains at the first contact point, producing a dark matter deficient gas clump (in red dotted circles in the middle column of Figure \ref{fig:9}). 
However, unlike our idealized high-resolution simulations in Sections \ref{sec:3} and \ref{sec:4}, the gas clump never turns into a galaxy, with only negligible star formation rate in the entire process (see the left panel of Figure \ref{fig:10}).  
The cumulative mass profile at $t=762$ Myr shows that the clump consists of mostly only gas --- $M_{\rm gas}=4.75\times10^{8}{\,\rm M_{\odot}}$ and $M_{\rm DM}=2.99\times10^{7}{\,\rm M_{\odot}}$ within 5 kpc (right panel of Figure \ref{fig:10}; note that most of the dark matter particles within 20 kpc are not dynamically associated with the clump). Since this gas clump is not self-gravitating, it will less likely be identified as “halos” by a halo finder. 
Indeed, in the bottom row of Figure \ref{fig:9} which displays only the particles that belong to the halos in the {\sc SubFind} catalogue, the gas particles in the aforementioned gas clump are almost entirely missing (in red dotted circles).\footnote{We have also visually inspected 8 of the 248 high-velocity collision events found in Section \ref{sec:4.1}, all of them in the main TNG100-1 volume, not in the ``subbox''.  Due to the coarse temporal resolution, however, we find it  very challenging to locate the sites of DMDG formation.  There are hints that some collisions produced dark matter deficient objects such as  gas clumps, but none of them could be considered as DMDGs.}  
Our study of the TNG100-1 snapshots themselves provides a clue on why it has been difficult to find  DMDGs in its halo catalogue (Section \ref{sec:5.1}).
We now discuss why such difference is observed between two simulations --- the idealized high-resolution simulations on {\sc Enzo} and {\sc Gadget-2} in Sections \ref{sec:3} and \ref{sec:4}, and the large cosmological simulation TNG100-1.  
\begin{figure}
    \centering
    \vspace{-1mm}    
    \includegraphics[width=0.47\textwidth]{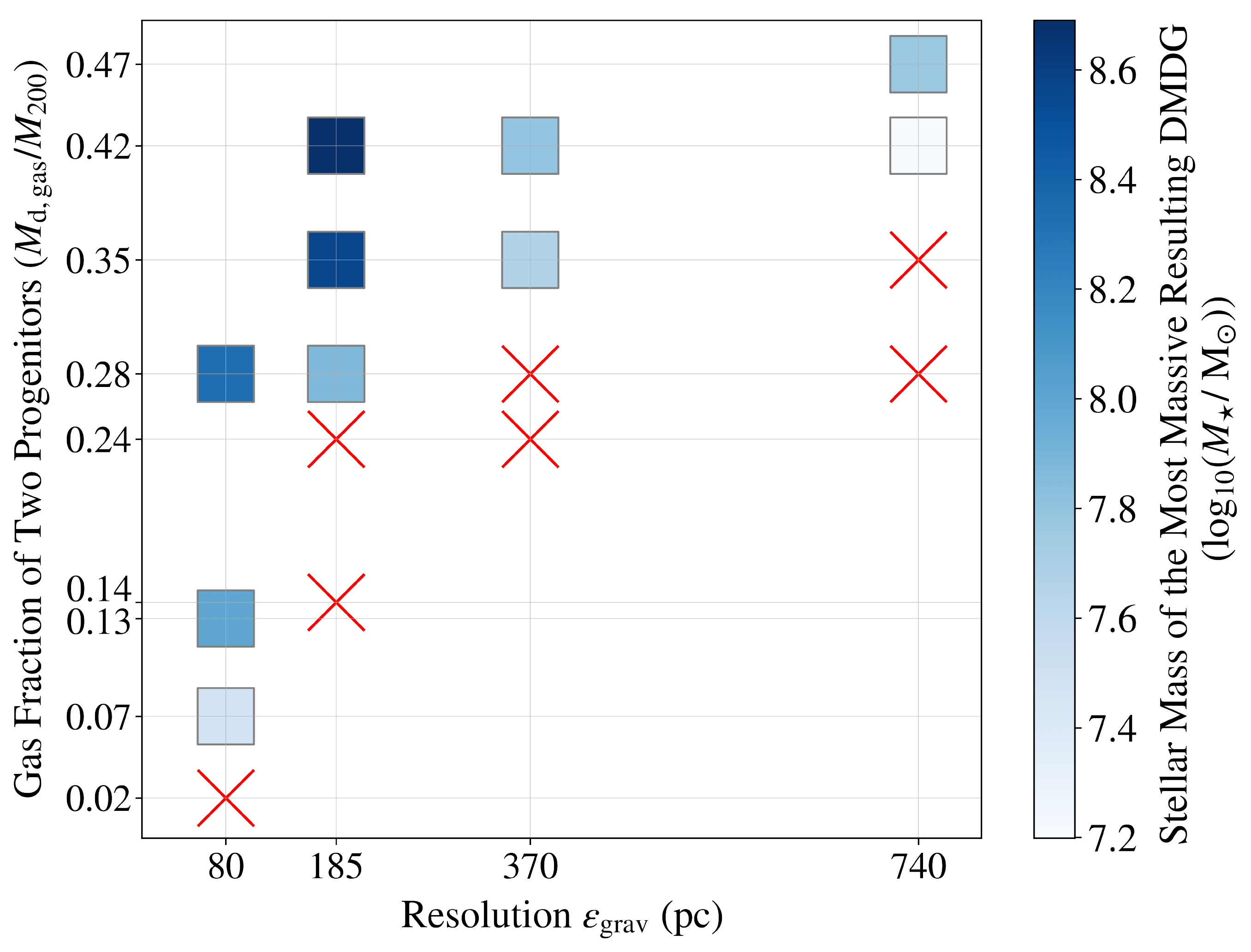}
    \vspace{-1mm}
    \caption{Results of idealized galaxy collision simulations on {\sc Gadget-2} tested with varying spatial resolutions ($\epsilon_{\rm grav}$) and initial gas fractions, as listed in Table \ref{tab:gadget-parameter-2}. 
    Both the star formation threshold density and the cooling floor in the lower-resolution runs ($>$ 80 pc) are chosen to match those in TNG100-1. 
    The face color of each datapoint displays the stellar mass in the most massive DMDG produced, while a {\it red cross} indicates that no DMDG has formed.  
    In the runs with spatial resolution $\epsilon_{\rm grav} \gtrsim $ 185 pc, DMDG formation appears to depend on the initial gas mass in the progenitors (or the initial gas fraction).  
    See Section \ref{sec:5.3} for more information.}
    \label{fig:13}
    \vspace{4mm}    
\end{figure}

\subsection{Difference Between High-resolution (80 pc) Simulations and the Large Cosmological Simulation: Importance of Spatial Resolution} \label{sec:5.3}

In this subsection, we discuss the reason why DMDGs do not form in the TNG100-1 universe (Sections \ref{sec:5.1}-\ref{sec:5.2}) but do form in the idealized galaxy collision simulations presented earlier (Sections \ref{sec:3.2}-\ref{sec:3.3}, and \ref{sec:4.3}). 
To illustrate the cause, we have performed a suite of resolution tests on {\sc Gadget-2} (see Table \ref{tab:gadget-parameter-2}). 
One of the tests, shown in Figure \ref{fig:11}, is identical to our fiducial run in Section \ref{sec:3.2} (Figure \ref{fig:3}) but with a lowered resolution --- the Plummer equivalent gravitational softening length of $\epsilon_{\rm grav}= 185$ pc,  and the minimum hydrodynamic smoothing length of $0.2 \epsilon_{\rm grav} = 37$ pc.  
These choices are to approximately match the softening achieved in the TNG100-1 run at $z=7$ --- a conservative way to represent {\sc IllustrisTNG}'s resolution near its highest (see Section \ref{sec:2.2} and footnote \ref{footnote:tng-res}). 
Particle mass resolution is accordingly lowered from Section \ref{sec:3.2} (Table \ref{tab:gadget-parameter}) to 
$m_{\rm DM}=4.26\times10^{6}{\,\rm M_{\odot}}$ and $m_{\rm \star, IC} = m_{\rm gas, IC}=1.40\times10^{6}{\,\rm M_{\odot}}$.  
We also adopt a lowered star formation threshold density $n_{\rm H, thres} = 0.13 \,{\rm cm^{-3}}$ and a cooling floor at $10^{4}$ K, again to match TNG100-1. 
All other collision parameters such as the collision velocity and disk angle are identical to those in its higher-resolution counterpart in Figure \ref{fig:3} (see also Table \ref{tab:gadget-parameter-2}), except the lowered gas fraction (from 0.28 to 0.14) to mimic the total gas mass of the progenitors in Figure \ref{fig:9} ($\sim 1.6 \times 10^{9}\,\rm M_{\odot}$).  

As in the higher-resolution simulations shown in Figures \ref{fig:2} and \ref{fig:4}, dark matter particles pass through one another while some gas particles interact and remain near the first contact surface (middle column of Figure \ref{fig:11}).
However, no galaxy --- DMDG or not --- is identified by  the {\sc Hop} algorithm in the intervening medium, in stark contrast to Figures \ref{fig:2} and \ref{fig:4}. 
The only new gravitationally bound structure above $10^{7}\,\rm M_{\odot}$ is the massive gas clump seen in the red dotted circles in the middle and right columns of  Figure \ref{fig:11}.  
The gas clump becomes diffused by $t=870$ Myr, left with little dark matter or stars. 
Only a few star particles are created around $t=0$ Myr with SFR $ < 0.1{\,\rm M_{\odot}}{\rm yr}^{-1}$ (left panel of Figure \ref{fig:12}).

Figure \ref{fig:13} summarizes the results of our extensive resolution study in Table \ref{tab:gadget-parameter-2}.  
The stellar mass of the most massive DMDG in each idealized galaxy collision simulation is shown in the plane of spatial resolution ($\epsilon_{\rm grav}$) and initial gas fraction ($f_{\rm gas}$). 
At fixed resolution, progenitors with higher gas fractions produce more massive collision-induced DMDGs, a trend already found in Section \ref{sec:3.3} and Table \ref{tab:gadget-parameter}. 
And at fixed gas fraction, simulations with higher resolution produce more massive DMDGs.  
In addition, we find that the minimum gas mass in the progenitors needed to induce any DMDG formation depends on the simulation resolution:  $2.92\times10^{8}\,\rm M_{\odot}$ (or $f_{\rm gas} > 0.02$) for $\epsilon_{\rm grav} =$ 80 pc, $2.81 \times 10^{9}\,\rm M_{\odot}$ (or $f_{\rm gas} > 0.24$)  for $\epsilon_{\rm grav} =$ 185 pc, and so forth. 
In other words, our findings imply that, at $z=7$ when the TNG100-1 run had a spatial resolution of 185 pc, only the collisions of very gas-rich galaxies ($f_{\rm gas} > 0.24$) would have had any chance of forming DMDGs.  

For the stricter criteria on the gas fraction of the progenitor galaxy ($f_{\rm gas} > 0.20$) in the findings in Section \ref{sec:4.1}, we find only four galaxy collision events. 
These collision events do not produce DMDGs because of consecutive re-merging of galaxy pairs shortly after the collision event or tidal strippings of the gas by one of the progenitors.
This poor population of the gas-rich dwarf galaxy is possibly due to the tidal stripping or ram-pressure while the satellites orbit their massive hosts \citep{2013MNRAS.432..336W}. Some may argue that numerical simulations may have mis-predicted the number of gas-rich dwarf galaxies which is often decided by the adopted feedback scheme (see e.g., Figure 4 of \cite{2018MNRAS.473.4077P}).

We suggest that the collision-induced DMDG formation is unlikely to occur in a (relatively) low-resolution simulation like TNG100-1 because {\it (1)} the multiphase medium in the intervening gas between the colliding galaxies --- and thus the tidally and shock- compressed high-density star-forming clumps --- is less likely to be resolved, and {\it (2)} the coarsely distributed gas particles may describe only a diffuse column of disk gas (i.e., small column density) that a gas particle in the other disk will clash into. We therefore argue that the collision-induced DMDGs are not found in the TNG100-1, neither in its halo catalogue nor in its snapshots themselves, due to the insufficient numerical resolution (and the dependence of gas mass required for DMDG formation on the resolution).
A sufficient numerical resolution that reliably resolves the multiphase ISM and the tidally and shock- compressed gas clumps appears to be critical to realizing the so-called collision-induced DMDG formation in simulations.  

\vspace{3mm}
\section{Conclusion} \label{sec:6}
\subsection{Summary} \label{sec:6.1}

Using gravito-hydrodynamics simulations, we have investigated the viability of the ``collision-induced'' formation scenario of a dark matter deficient galaxy (DMDG) for the first time.  
With a suite of idealized high-resolution (80 pc) galaxy collision simulations on both mesh-based and particle-based codes, we find that the recently observed DMDGs (e.g., NGC1052-DF2 and NGC1052-DF4) could have formed when gas-rich, dwarf-sized galaxies collide with a high relative velocity of $\sim 300\,{\rm km\,s^{-1}}$ (Sections \ref{sec:3.1}-\ref{sec:3.2}).
The difference in the nature of dark matter and baryon separates the two components in such a supersonic collision. 
The warm disk gas dissociated from dark matter is compressed by shock and tidal interaction to form stars, subsequently a DMDG.   
We then explore the parameter space of the collision such as the relative velocity, disk angle, and gas fraction, and determine  that one of the important factors for DMDG formation is the amount of gas that actually participates in the  collision (Section \ref{sec:3.3}). 

Then, by inspecting 95 snapshots from $z=10$ to 0.01 of the large cosmological simulation TNG100-1, we identify 248 high-velocity galaxy collision events  in which collision-induced DMDGs are expected to form (Section \ref{sec:4.1}).  
Adopting a representative collision configuration from one of these events in a high-resolution (80 pc) numerical experiment, we show that a long-lasting DMDG can form near a massive host galaxy when two gas-rich, dwarf satellites collide with a high relative velocity (Section \ref{sec:4.3}).  
However, we have not found any evidence that the small number of DMDGs in the TNG100-1 halo catalogue are related to the high-velocity collision of nearby galaxies (Section \ref{sec:5.1}).
We instead reveal that in the {\sc TNG100-1-Subbox0} volume, at least one galaxy collision event produced a giant gas clump with little dark matter.  
Yet, unlike our idealized high-resolution simulations, the gas clump neither turns into a galaxy, nor is identified as a halo in the TNG100-1 halo catalogue, explaining why it was difficult to locate a collision-induced DMDG in the catalogue (Section \ref{sec:5.2}).  
With an extensive resolution study, we argue that the resolution of TNG100-1 ($\epsilon_{\rm grav}^{z=0} = 740\, {\rm pc}$ for collisionless particles) is likely insufficient to reproduce the DMDG formation during galaxy collisions (Section \ref{sec:5.3}).  

\subsection{Future Work} \label{sec:6.2}

Our results demonstrate a unique path in which a galaxy could form with an unconventional dark matter content.  
As we try to test the new ``collision-induced'' formation mechanism of DMDGs, interesting ideas for future projects are being actively explored to expand the scope of our study.  
\begin{itemize}
\item  As discussed in Section \ref{sec:3.2}, the limited numerical resolution in our idealized experiments prevents us from resolving the internal dynamics and the detailed morphology of the resulting DMDGs.
A pair of colliding galaxies resolved with much higher resolution ($\lesssim$ 1-10 pc) may allow us to directly compare the simulated DMDGs with NGC1052-DF2 and NGC1052-DF4. 
The higher-resolution simulation could be used to check if the collision-induced DMDG formation scenario better explains the observed characteristics of NGC1052-DF2 and NGC1052-DF4 (e.g., number of bright globular clusters, effective radius, surface brightness; see Section \ref{sec:1}) and reveals the main mechanism of star formation in the collision-induced DMDGs.
\item The collision-induced galaxy formation scenario discussed in the present article is reminiscent of the recently studied mechanism for globular cluster formation in merging proto-galaxies and substructures \citep[e.g.,][]{2018MNRAS.474.4232K, 2020ApJ...890...18M}.  
The higher-resolution simulations like the one discussed above may reveal that the high-velocity collision of structures grants a unique opportunity to form objects that are with little dark matter, a primary driver of gravitational collapse in most structure formation.    
Depending on the sizes of the colliding progenitors, the resulting ``collision-induced'' objects could be anything from globular clusters to DMDGs.
\item Known as the diversity problem of dwarf galaxy rotation curves, some dwarf galaxies exhibit very different density profiles than most others, often indicating a net removal of a large amount of mass from the inner region \citep[e.g.,][]{2015MNRAS.452.3650O, 2018MNRAS.473.4392S,2019arXiv191109116S}. 
Dwarf-sized galaxies that have undergone a high-velocity collision could have lost a significant amount of mass from its inner region. 
We are actively investigating if such collisions could explain not only the removal of baryons from a dark matter halo, but also the {\it partial} removal of dark matter from the halo's center, thereby resolving the diversity problem of rotation curves.  
\end{itemize}

\section*{Acknowledgments}

The authors thank Ena Choi, J\'{e}r\'{e}my Fensch, Seung-o Ha, Taysun Kimm, Choonkyu Lee, Oliver M\"{u}ller and Joseph Silk for insightful discussion during the progress of this study.
The authors also thank Dylan Nelson and Vicente Rodriguez-Gomez for providing {\sc SubLink-Gal}, the baryon-based merger trees for the {\sc IllustrisTNG} simulation.
Ji-hoon Kim acknowledges support by Samsung Science and Technology Foundation under Project Number
SSTF-BA1802-04, and by Research Start-up Fund for the new faculty of Seoul National University.
This work was also supported by the National Institute of Supercomputing and Network/Korea Institute of Science and Technology Information with supercomputing resources including technical support, grants KSC-2018-CRE-0052 and KSC-2019-CRE-0163. 
The publicly available {\sc Enzo} and {\tt yt} codes used in this work are the products of collaborative efforts by many independent scientists from numerous institutions around the world. Their commitment to open science has helped make this work possible.

\bibliography{cites}{}
\bibliographystyle{apj1}

\label{lastpage}

\end{document}